\documentclass[format=acmsmall, review=false]{acmart}
\makeatletter
\def\@ACM@checkaffil{% Only warnings
    \if@ACM@instpresent\else
    \ClassWarningNoLine{\@classname}{No institution present for an affiliation}%
    \fi
    \if@ACM@citypresent\else
    \ClassWarningNoLine{\@classname}{No city present for an affiliation}%
    \fi
    \if@ACM@countrypresent\else
        \ClassWarningNoLine{\@classname}{No country present for an affiliation}%
    \fi
}
\makeatother

\usepackage{acm-ec-25}
\usepackage{booktabs} % For formal tables
\usepackage[ruled]{algorithm2e} % For algorithms

\usepackage{algpseudocode}
\usepackage{tikz}
\usepackage{amsmath}
\usepackage{amssymb}
\usepackage{bbm}
\usepackage{rotating}
\usepackage{fancyvrb}
\usepackage{float} % 在导言区引入 float 包
\usepackage{anyfontsize}
\usepackage{tabularx} % 引入tabularx包
\usepackage{booktabs} % 用于提高表格的视觉效果
\usepackage{graphicx}
\usepackage{threeparttable} % 用于创建带注脚的表格
\usepackage{caption}
\usepackage[labelfont=sf]{subcaption}
\usepackage{cleveref}
\usepackage{placeins}
\usepackage{multirow}
\usepackage{natbib}
 \bibpunct[, ]{(}{)}{,}{a}{}{,}%

\SetAlFnt{\small}
\SetAlCapFnt{\small}
\SetAlCapNameFnt{\small}
\SetAlCapHSkip{0pt}
\IncMargin{-\parindent}

\setcitestyle{authoryear}

\title[Assessing Uncertainty in Stock Returns: A Gaussian Mixture Distribution-Based Method]{Assessing Uncertainty in Stock Returns: A Gaussian Mixture Distribution-Based Method}

\author{Yanlong Wang}
\affiliation{
  \institution{Tsinghua University}
  \country{China}
}
\email{wangyanl21@mails.tsinghua.edu.cn}

\author{Jian Xu}
\affiliation{
  \institution{Tsinghua University}
  \country{China}
}
\email{xujian20@mails.tsinghua.edu.cn}

\author{Shao-Lun Huang}
\affiliation{
  \institution{Tsinghua University}
  \country{China}
}
\email{shaolun.huang@sz.tsinghua.edu.cn}

\author{Danny Dongning Sun\textsuperscript{\raise-0.9ex\hbox{*}}}
\affiliation{
  \institution{Peng Cheng Laboratory}
  \country{China}
}
\email{ds316@columbia.edu}

\author{Xiao-Ping Zhang}
\authornote{Corresponding Authors}
\affiliation{
  \institution{Tsinghua University}
  \country{China}
}
\email{xiaoping.zhang@sz.tsinghua.edu.cn}
\begin{abstract}
This study seeks to advance the understanding and prediction of stock market return uncertainty through the application of advanced deep learning techniques. We introduce a novel deep learning model that utilizes a Gaussian mixture distribution to capture the complex, time-varying nature of asset return distributions in the Chinese stock market. By incorporating the Gaussian mixture distribution, our approach effectively characterizes short-term fluctuations and non-traditional features of stock returns—such as skewness and heavy tails—that are often overlooked by traditional models. Compared to GARCH models and their variants, our method demonstrates superior performance in volatility estimation, particularly during periods of heightened market volatility. It provides more accurate volatility forecasts and offers unique risk insights for different assets, thereby deepening the understanding of return uncertainty. Additionally, we propose a novel use of Code embedding which utilizes a bag-of-words approach to train hidden representations of stock codes and transforms the uncertainty attributes of stocks into high-dimensional vectors. These vectors are subsequently reduced to two dimensions, allowing the observation of similarity among different stocks. This visualization facilitates the identification of asset clusters with similar risk profiles, offering valuable insights for portfolio management and risk mitigation. Since we predict the uncertainty of returns by estimating their latent distribution, it is challenging to evaluate the return distribution when the true distribution is unobservable. However, we can measure it through the CRPS (Continuous Ranked Probability Score) to assess how well the predicted distribution matches the true returns, and through MSE (Mean Squared Error) and QLIKE (quasi-likelihood) metrics to evaluate the error between the volatility level of the predicted distribution and proxy measures of true volatility. The results highlight the robustness and predictive accuracy of our approach, suggesting that it significantly improves volatility forecasting and enhances the practical utility of risk modeling in financial markets. In summary, we assess return uncertainty at each timestep, performing extensive experiments across Chinese equities and measuring uncertainty similarities among stocks. With the Transformer-based model enhanced by stock ticker embeddings, our framework outputs the Gaussian mixture distribution to characterize the return uncertainty, offering a new approach to study return volatility.
\end{abstract}

\begin{document}

% Title page for title and abstract only.
\begin{titlepage}

\maketitle
\clearpage
% Optionally include a table of contents
\vspace{-0.4em}
\setcounter{tocdepth}{2} % adjust to 1 if desired
\tableofcontents

\end{titlepage}

% Paper body
\section{Introduction}
\label{sec:Introduction}

Accurate forecasting of stock market returns and their uncertainty is critically important for financial risk management and optimization of investment decisions. Predicting stock returns is challenging and inherently uncertain, making it crucial to account for this uncertainty. The potential distribution of forecasted returns is a great representation of this uncertain nature. 
% However, recognizing the dynamic distribution of returns is challenging for the following reasons: 1) the actual dynamic distribution of returns is unknown, 2) it is difficult to present and measure their complex distribution variations, and 3) it is hard to utilize the nonlinear relationships in information flow. Consequently, single distributions, such as the t-distribution or normal distribution, have been predominantly used to describe the return distribution and using statistical indicators to measure its risk.
However, recognizing the dynamic distribution of returns presents several challenges: 1) the true dynamic distribution of returns is unknown, 2) capturing and measuring its complex variations is difficult, and 3) leveraging the nonlinear relationships in information flow is challenging. As a result, traditional approaches often rely on single distributions, such as the t-distribution or normal distribution, to model return distributions and use statistical indicators to measure risk.

The volatility of asset prices is a key measure of risk and uncertainty in evolving financial markets~\citep{ross1989information}. Early studies, such as the mixture distribution hypothesis~\citep{clark_subordinated_1973}, highlighted the strong link between asset price volatility and the cumulative impact of information flow. This theory explains how market prices respond to new information, suggesting that asset price movements are not simply random fluctuations but are closely tied to how market participants interpret and react to information.
Building on this, \cite{darolles_mixture_2017} expanded on the idea that in liquidity-constrained market environments, asset price changes may not immediately reflect all available information. This concept aligns with the sequential information arrival hypothesis~\citep{copeland_model_1976, easley_order_1991, easley_time_1992}, which describes how information disseminates slowly across the market as different types of investors gradually receive and process it. \cite{Han2009} further argues that stock price momentum can arise from uncertainty about the accuracy of cash flow forecasts, with momentum reflecting investors' evolving understanding of the relative reliability of information sources and their subsequent updates to expectations.
Several studies, including \cite{Ang2006}, \cite{ENGLE1993}, \cite{GLOSTEN1993}, and \cite{SCHWERT1990}, focus on the relationship between price changes and information flow. The most widely used models for capturing volatility dynamics in financial time series are GARCH-type models. A wealth of research~\citep[e.g.][]{bollerslev_generalized_1986, nelson_1991, engle1982, ZAKOIAN1994931, ding_long_1993, baillie_frac_1996} has demonstrated that the volatility aggregation phenomenon in time series data can be effectively captured by GARCH models and their variants, making them valuable tools for understanding and forecasting financial market volatility. However, these models rely on strong prior assumptions and are limited in their ability to capture complex nonlinear patterns, especially when dealing with multivariate and multi-asset data~\citep{engle1995multivariate, engle2002dynamic, franses_additive_1999, Andersen2003, Kenneth1998}.

% Machine learning has the capability to learn complex nonlinear relationships within data, allowing it to capture intricate patterns and dependencies that may not be easily captured by traditional linear models or statistical methods~\citep{gu2020empirical,LEIPPOLD202264}. Many research~\citep[e.g.][]{zhou2024much, Bianchi2020, ban2018machine, avramov2023machine, huang2023machine} highlights that machine learning techniques outperform traditional methods in predicting asset returns, and this superiority extends across various asset types including stocks, bonds, funds, and portfolio management strategies. Deep learning, as an advanced machine learning approach outperforms simple machine learning methods due to its stronger learning capabilities and better handling of complex data structures and patterns~\citep{gu2020empirical, LEIPPOLD202264, chen2024deep}. Common deep learning methods include Neural Networks, Long Short-Term Memory networks (LSTM), Convolutional Neural Networks (CNN) and their varients. These methods are widely used in asset pricing. Additionally, Transformers~\citep{vaswani2017attention}, an emerging method, are increasingly being applied in asset management and other management science fields due to their excellent performance in handling sequential data~\citep{Soohan2024, Álvaro2024, puranam2021impact}. 

Machine learning is capable of learning complex nonlinear relationships within data, enabling it to capture intricate patterns and dependencies that may be difficult for traditional linear models or statistical methods to uncover~\citep{gu2020empirical, LEIPPOLD202264}. Numerous studies~\citep[e.g.][]{zhou2024much, Bianchi2020, ban2018machine, avramov2023machine, huang2023machine} highlight that machine learning techniques often outperform traditional methods in predicting asset returns, demonstrating superior performance across various asset classes, including stocks, bonds, funds, and portfolio management strategies. Among machine learning methods, deep learning stands out due to its enhanced learning capabilities and superior ability to handle complex data structures and patterns~\citep{gu2020empirical, LEIPPOLD202264, chen2024deep}. Common deep learning techniques, such as Neural Networks, Long Short-Term Memory (LSTM) networks, and Convolutional Neural Networks (CNN) and their variants, are widely applied in asset pricing. Additionally, Transformers~\citep{vaswani2017attention}, an emerging approach, are gaining traction in asset management and other fields of management science due to their excellent performance in processing sequential data~\citep{Soohan2024, Álvaro2024, puranam2021impact}.

% In terms of uncertainty, different researchers adopt various methods and measures. \cite{bali2016risk} uses the variance risk premium as a proxy for uncertainty. \cite{anderson2009impact} measures uncertainty through the degree of disagreement among professional forecasters. \cite{baltussen2018unknown} adopts that the volatility of volatility is a measure of uncertainty. \cite{abdar2021review} views uncertainty quantification, in a wide range of fields, as the process of evaluating the potential distribution of model outputs based on existing information, and reviews uncertainty quantification solutions using deep learning methods.

In terms of uncertainty, various methods and measures are adopted to assess uncertainty. For instance, \cite{bali2016risk} uses the variance risk premium as a proxy for uncertainty, while \cite{anderson2009impact} measures uncertainty through the degree of disagreement among professional forecasters. \cite{baltussen2018unknown} suggests that the volatility of volatility serves as a measure of uncertainty. Additionally, \cite{abdar2021review} defines uncertainty quantification—across a wide range of fields—as the process of evaluating the potential distribution of model outputs based on existing information, and reviews deep learning-based solutions for uncertainty quantification.

\begin{figure}
    \centering
    \includegraphics[width=1.0\textwidth]{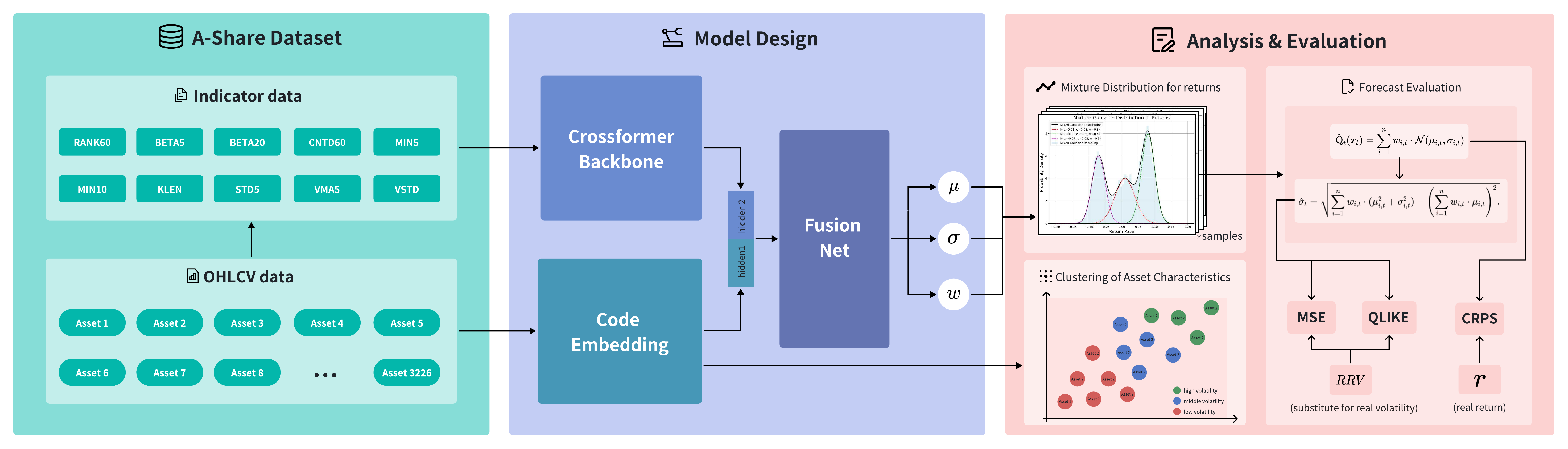}
    \caption{Overall Procedure. During the dataset phase, we collect historical price  and volume data for over 3000 A-share stocks and construct a series of  indicator data to serve as input for the model. In the model design  phase, the Crossformer structure is used to process time-series indicators, and Code Embedding is employed to represent stock tickers, these representations are then fed into fusion Net to predict potential return distributions and analyze asset attributes. In the analysis and evaluation phase, a series of evaluation metrics and visualization methods are used to test the model's performance.}
    \label{fig:net_structure}
\end{figure}

In this work, we aim to utilize deep learning technique to predict the dynamic distribution of asset price changes and to explore the nonlinear relationship between information flow and price volatility. Specifically, we represent the distribution of the asset return by a Gaussian mixture distribution\footnote{Also known as the mixture of Gaussian distributions.} and predict its parameters in the output layer of the deep learning network. Compared with the traditional single distributions, such as Gaussian or t-distribution, which fail to precisely capture the specific return distribution in each give period~\citep{han2022bimodal, boll_quasi_1992}, the Gaussian mixture distribution has the capacity to capture a wide range of distributions. The mixture distribution methodology has been demonstrated the ability for asset pricing~\citep{rachev1993laplace}, and can be optimized by maximum likelihood estimation~\citep{jagabathula2020conditional, Nicholas1982}. However, due to computational constraints, underdeveloped machine learning techniques, and optimization challenges associated with mixture distributions, prior research failed to achieve substantial progress in this domain.

% Our deep learning model is particularly designed to suit for the needs of dealing with time series data and differentiating the stock identities. Firstly, we use the Crossformer architecture, which is a variant of the Transformer, to process indicator data. This architecture includes several modules designed to improve its capability in handling sequential information~\citep{zhang_crossformer_2023}. Secondly, in dealing with multiple assets and effectively identifying the particularity of each asset, we construct a code embedding network to encode the stock codes into a vector space. Finally, we build a fusion network. The outputs from the Crossformer and the code embedding network are merged and used as the input to the fusion network. The fusion network outputs the parameter values of the Gaussian mixture distribution and constructs the prediction of the asset return distribution. We called this model as MDNe, and present the total network structure in~\Cref{fig:net_structure}. Besides, we also build a model without code embedding part and called it as MDN. To observe the MDNe model's understanding of the level of uncertainty in stock returns, we use the t-SNE dimension-reduction technique to decrease the outputs of the code embedding part to two-dimensional space~\citep{maaten_visualizing_2008}, and visualize the clustering of the uncertainty features among stocks.

% {\color{red} relation between those two paragraphs, how to change the topic from mixture distribution to deep learning model.}

Our proposed deep learning model is specifically designed to handle time series data and distinguish between individual stock identities. First, we utilize the Crossformer architecture, a variant of the Transformer, to process indicator data. This architecture includes several modules that enhance its ability to handle sequential information~\citep{zhang_crossformer_2023}. Next, to effectively manage multiple assets and capture the unique characteristics of each stock, we construct a code embedding network that encodes stock identifiers into a vector space. Finally, we build a fusion network that combines the outputs of both the Crossformer and the code embedding network, which are then used as input to generate predictions. The fusion network outputs the parameter values for a Gaussian mixture distribution, constructing the predicted asset return distribution. We refer to this model as MDNe and present its full network structure in~\Cref{fig:net_structure}. Additionally, we build a version of the model without the code embedding component, referred to as MDN.
To evaluate MDNe's understanding of the uncertainty in stock returns, we apply the t-SNE dimension-reduction technique to map the outputs of the code embedding part to two-dimensional space~\citep{maaten_visualizing_2008}. This allows us to visualize the clustering of uncertainty features across stocks.

Our empirical analysis is based on a dataset comprising all stocks listed on the main board of the Chinese A-share market from 2018 to 2022. The dataset includes daily stock prices (open, close, high, and low), trading volumes, and 10 indicators that capture changes in information flow (see \Cref{tb:summary_indicators}). Additionally, daily realized range volatility is computed using 5-minute intraday price data. The inclusion of such a large number of stocks ensures the model’s generalizability and strengthens the robustness of our results. To evaluate the performance of our model, we use several standard metrics, including Mean Squared Error (MSE), Mean Absolute Error (MAE), and Continuous Ranked Probability Score (CRPS), and compare it against other volatility models, such as GARCH and its variants.

% To the best of our knowledge, this work is the first to integrate deep learning techniques with the Gaussian mixture distribution model to analyze asset return uncertainty. We show that this novel method outperforms the GARCH model and its variants in volatility forecasting. By encoding individual asset characteristics and utilizing embedding techniques, we uncover insights into the similarities in return uncertainty across different assets. This research introduces a pioneering deep learning-based approach that combines Gaussian mixture distributions for forecasting the time-varying distribution of asset returns, offering fresh perspectives for risk modeling. Furthermore, by applying embedding techniques and the t-SNE method to map risk attributes from high-dimensional to two-dimensional space, we visualize the clustering of risk characteristics among assets, providing valuable insights into their relationships.

\begin{itemize}
    \item \textbf{Development of a Dedicated Deep Learning Framework}. Construct a Transformer-based variant deep learning model to process time-series input data, integrated with stock code embedding techniques for encoding symbol-specific information. The hidden representations from both modalities are fused through a neural network, which ultimately generate a Gaussian mixture distribution to predict dynamic return distributions of stocks. This approach effectively captures complex characteristics and nonlinear relationships in stock returns, addressing the limitations of conventional models.

    \item \textbf{Construction of Stock Code Embedding and Visualization Methodology}. Convert stock uncertainty features into high-dimensional vectors using bag-of-words modeling, followed by t-SNE dimensionality reduction for visual exploration. This framework reveals similarities in uncertainty patterns across different assets, offering novel tools and perspectives for investigating inter-asset relationships, while enabling deeper insights into shared uncertainty characteristics through embedding analysis.

    \item \textbf{Innovative Risk Modeling Perspective and Empirical Validation}. Implement stepwise uncertainty assessment for temporal return sequences, establishing a new paradigm for risk modeling. Empirical comparisons with traditional methods like GARCH demonstrate superior volatility forecasting performance, enhancing practical risk evaluation accuracy for investment decisions. This work pioneers a methodology combining deep learning with Gaussian mixture distributions to predict time-varying asset return distributions, opening new research directions for advanced risk modeling.
\end{itemize}

% {\color{red} where to insert a brief literature review / related work, could the second and third paragraphs be extracted as the related work section?}
% The remainder of the paper is organized as follows. We introduce data and methods in \Cref{sec:methods}, present the empirical results in \Cref{sec:empirical results}, then conduct the robustness analysis in \Cref{sec:robustness analysis}. Finally, our conclusion is found in \Cref{sec:conclusion}.

\section{Dataset and Method}
\label{sec:methods}

\subsection{Data Source}
\label{sec:data}

We collected data for all A-share main board listed companies in China from January 2018 to December 2022. The dataset was divided as follows: data samples from 2018 to 2020 were used as the training set to optimize the model parameters, while data samples from 2021 to 2022 served as the test set to verify the out-of-sample performance of the trained model. Additionally, for the deep learning model, we selected six-month data samples from the second half of 2020 as the validation set to adjust hyperparameters and prevent overfitting. Since the GARCH model and its variants have fewer hyperparameters, no separate validation set was used for them.

To meet the requirements of GARCH models, which need sufficient sample points for each stock, we excluded stocks listed or delisted in the past two years, resulting in a final dataset of 3226 stock samples. For the deep learning model, we constructed 10 distinct technical indicators based on daily trading volume and price data from these stocks. The detailed formulas for these indicators can be found in \Cref{tb:summary_indicators}, which capture information from both volume and price data across time and cross-sectional dimensions.

\renewcommand{\arraystretch}{1.3}
\normalsize
\begin{table}[tbp]
\centering
\caption{Summary of indicators.}
\label{tb:summary_indicators}
\small
\begin{tabularx}{1\textwidth}{l l X} 

\hline
Indicator & Formula & Description \\
\cline{1-3}
RANK60 & Rank(close, 60) & Get the percentile of current close price in past 60 day's close price. \\

BETA5, &  \multirow{2}{*}{Slope(close, 5 or 20)/close}  & \multirow{2}{\linewidth}{The rate of close price change in the past 5 or 20 days, divided by latest close price.} \\
BETA20 & & \\

\multirow{3}{*}{CNTD60} & UP60 = Mean(close \textgreater preclose, 60) & \multirow{3}{\linewidth}{The difference between the percentage of days with price increases and the percentage of days with price decreases over the past 60 days.}\\
 & DOWN60 = Mean(close \textless preclose, 60) & \\
 & CNTD60 = UP60 - DOWN60 & \\
 
MIN5, & \multirow{2}{*}{Min(low, 5 or 10)/close}  & \multirow{2}{\linewidth}{The lowest price for past 5 or 10 days, divided by latest close price to remove unit.}\\
MIN10 & & \\

KLEN &  (high - low)/open  & The difference between the highest and lowest prices divided by the opening price yields the price volatility ratio\\

STD5 & Std(close, 5)/close & The standard diviation of close price for the past 5 days, divided by latest close price to remove unit \\

VMA5 & Mean(volume, 5)/volume & The mean value of volume for the past 5 days divided by lastest volume to remove unit\\

VSTD10 & Std(volume, 10)/volume & The standard deviation for volume in past 10 days divided by lastest volume\\
\hline
\end{tabularx}
\end{table}

In addition, we used intraday data with a 5-minute frequency to compute the realized range volatility (RRV) as described in \Cref{eq:rrv}. 
\begin{equation}
    \label{eq:rrv}
    \mathrm{RRV}_t = \frac{1}{4\log(2)} \sum_{i=1}^{n} ( \log(\mathrm{H}_{i,t}) - \log(\mathrm{L}_{i,t}) )^2,
\end{equation}
where $\mathrm{H}_{i,t}$ and $\mathrm{L}_{i,t}$ respectively present the highest and lowest prices observed in interval $i=1,...,n$ on day $t$. 
The RRV serves as a reliable proxy for the actual volatility, which actually cannot be directly measured. Specifically, the RRV represents the cumulative volatility derived from the amplitude within each time interval throughout the day. Previous research by~\cite{CHRISTENSEN2007323} demonstrates the proximity of the RRV to actual volatility, affirming its validity as a substitute.

\subsection{Data Preprocessing}
\label{sec:Data preprocessing}
We employ various preprocessing methods, including Z-Score normalization, depolarization, and handling missing values. Specifically, when removing outliers, we apply a threshold at the 1st and 99th percentiles to avoid losing sensitivity to abnormal values. This approach limits the data values to the numerical range between these percentiles, setting any values outside this range to the corresponding boundary value.

Regarding missing value imputation, a common method for input data in deep learning is to fill missing values with zeros. This approach helps avoid introducing bias and simplifies the processing procedure, as the standardized data values often cluster near zero. However, for indicator data, forward filling is more common, where the value from the previous period is used to fill the missing value in the current period. This strategy maintains consistency of features in the time dimension.

Factor standardization employs the Z-score standardization method, which applies Z-score operations to the indicator data of different stocks in the cross-sectional dimension. This processing assists in assessing the relative performance or characteristic values of each stock at a given point in time, while eliminating the influence of data magnitude and unit, ensuring comparability and fairness of indicator values among different stocks. Its formula is expressed as follows:
\begin{equation}
    \label{eq:zscore}
    \mathrm{Z} = \frac{\mathrm{X} - \mu}{\sigma}.
\end{equation}
For each cross-sections, $\mathrm{X}$ represents the original indicator vector value, $\mu$ indicates the mean value of an indicator across all stocks, $\sigma$ represents the standard deviation value of an indicator across all stocks, and $\mathrm{Z}$ is the value after Z-score standardization.

\subsection{Gaussian Mixture Distribution}
\label{sec:Gaussian mixture distribution}
It is well known that individual stock return distributions in general deviate from normal or t-distributions, exhibiting more complex patterns such as bimodality or skewness. The Gaussian mixture distribution is a powerful statistical tool that accurately captures and represents the shape of data distributions by combining multiple Gaussian distributions, making it well-suited for characterizing complex distributions of financial asset returns. Its effectiveness is supported by~\cite{giacomini_mixtures_2008}, who concluded that the mixture distribution can effectively approximate various types of return distributions. In this paper, we adopt the Gaussian mixture distribution as the model for the stock returns.

The variable \(x_t \in \mathbb{R}\) represents the space of return values at time step \(t\). The basic mathematical form of a Gaussian mixture distribution \(\mathrm{Q}_t(x_t)\), which models the distribution of returns, is a weighted sum of multiple Gaussian distributions as expressed on Equation \eqref{eq:gmm_time_index}. Each Gaussian distribution is defined by its mean \(\mu_{i,t}\), standard deviation \(\sigma_{i,t}\), and is weighted by \(w_{i,t}\) to form \(\mathrm{Q}_t(x_t)\). These weights sum to 1, ensuring that the entire mixture distribution is a valid representation of the probability distribution. Specifically, the mixture distribution takes the form:
\begin{equation}
\label{eq:gmm_time_index}
\begin{gathered}
\mathrm{Q}_t(x_t) = \sum_{i=1}^{n} w_{i,t} \cdot \mathcal{N}(\mu_{i,t}, \sigma_{i,t}),\\
\sum_{i=1}^{n} w_{i,t} = 1,
\end{gathered}
\end{equation}
where \(t = 1, 2, 3, \ldots, m\) represents the time steps, \(\mathrm{Q}_t(x_t)\) represents the distribution of returns at time \(t\), \(x_t\) represents the return value, \(w_{i,t}\) represents the weight of the \(i\)-th Gaussian component, \(\mu_{i,t}\) represents the mean of the \(i\)-th Gaussian component, and \(\sigma_{i,t}\) represents the standard deviation of the \(i\)-th Gaussian component. The parameter \(n\) represents the number of Gaussian components. 

% Consider a Gaussian mixture distribution composed of five one-dimensional Gaussian distributions as an example. In \Cref{fig:grid_of_images}, the dotted lines of various colors represent the individual Gaussian distributions, while the solid line represents the probability density function of the Gaussian mixture distribution composed of these components. It's evident that the weighted sum of these components with specific weights yields a uniquely determined Gaussian mixture distribution. With different sets of weights, the Gaussian mixture distribution exhibits a variety of non-traditional distribution patterns.

\begin{figure}
    \centering
    \begin{subfigure}[b]{0.4\linewidth}
        \captionsetup{font=small}  % 局部设置字体大小
        \includegraphics[width=\linewidth]{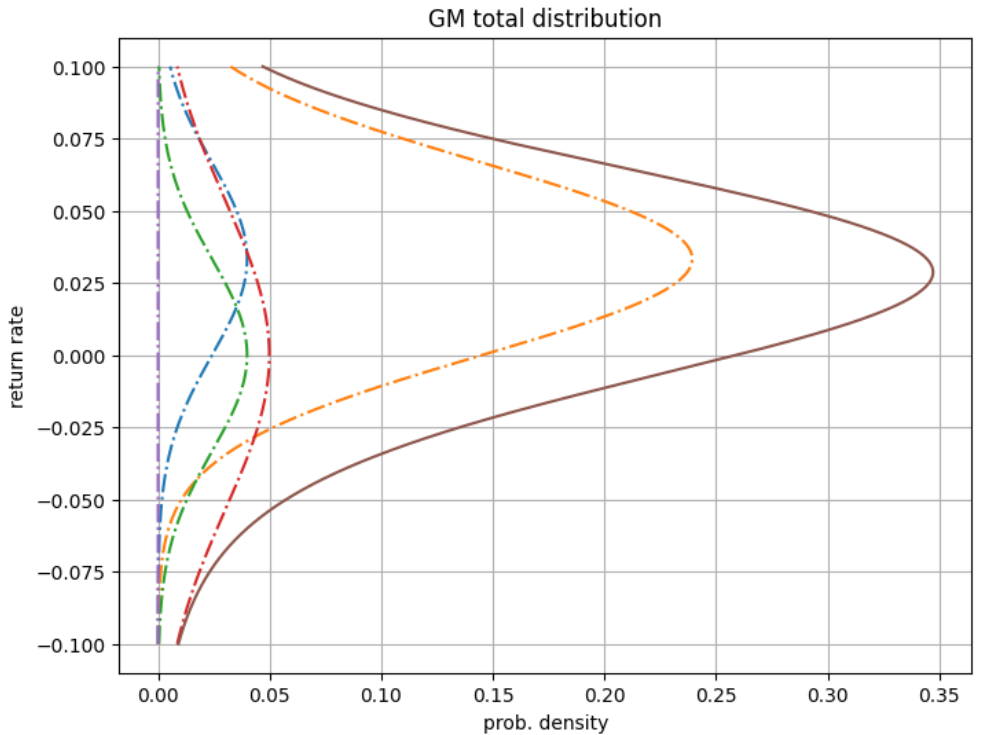}
        \caption{Normal distribution}
    \end{subfigure}
    \begin{subfigure}[b]{0.4\linewidth}
        \captionsetup{font=small}  % 局部设置字体大小
        \includegraphics[width=\linewidth]{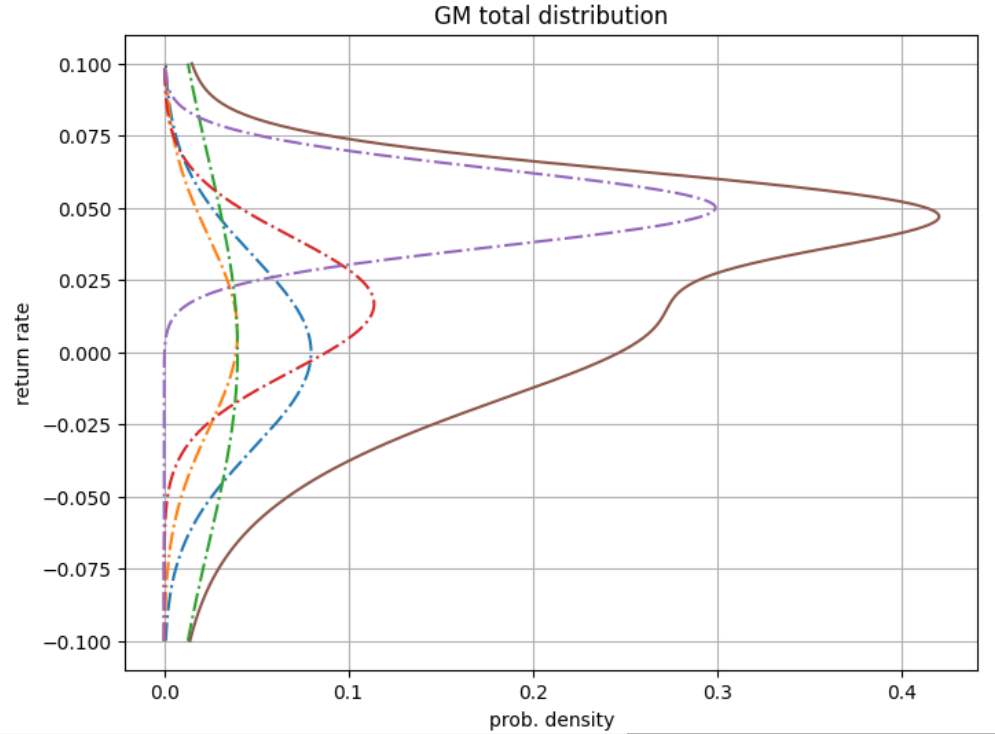}
        \caption{Skewed distribution}
    \end{subfigure}
    \begin{subfigure}[b]{0.4\linewidth}
        \captionsetup{font=small}  % 局部设置字体大小
        \includegraphics[width=\linewidth]{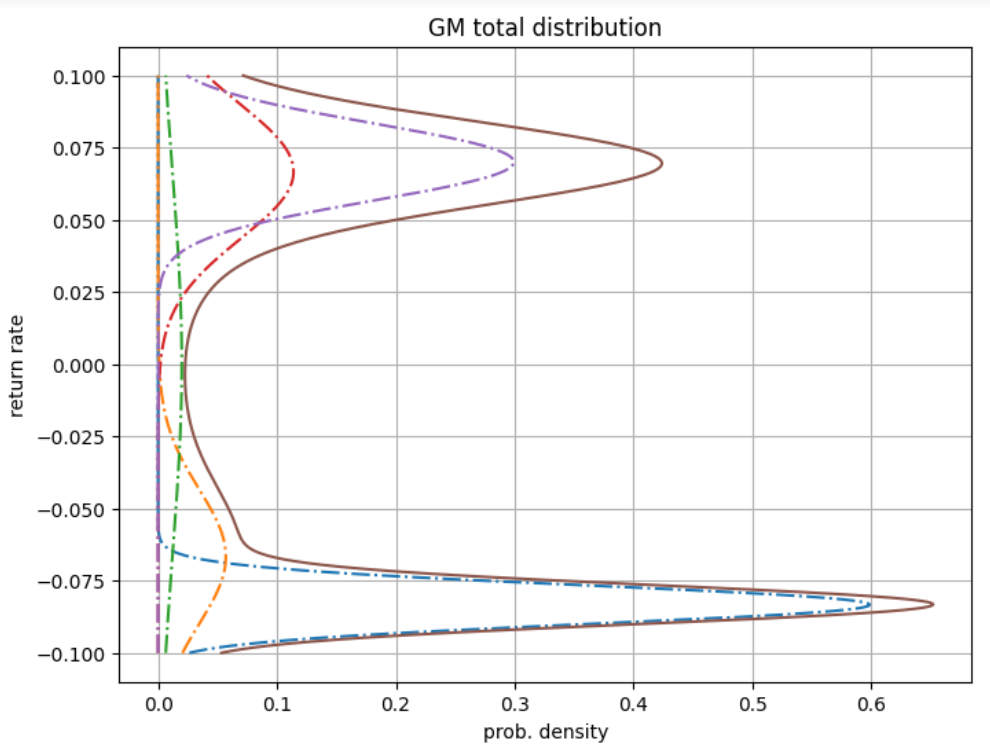}
        \caption{Bimodal distribution}
    \end{subfigure}
    \begin{subfigure}[b]{0.4\linewidth}
        \captionsetup{font=small}  % 局部设置字体大小
        \includegraphics[width=\linewidth]{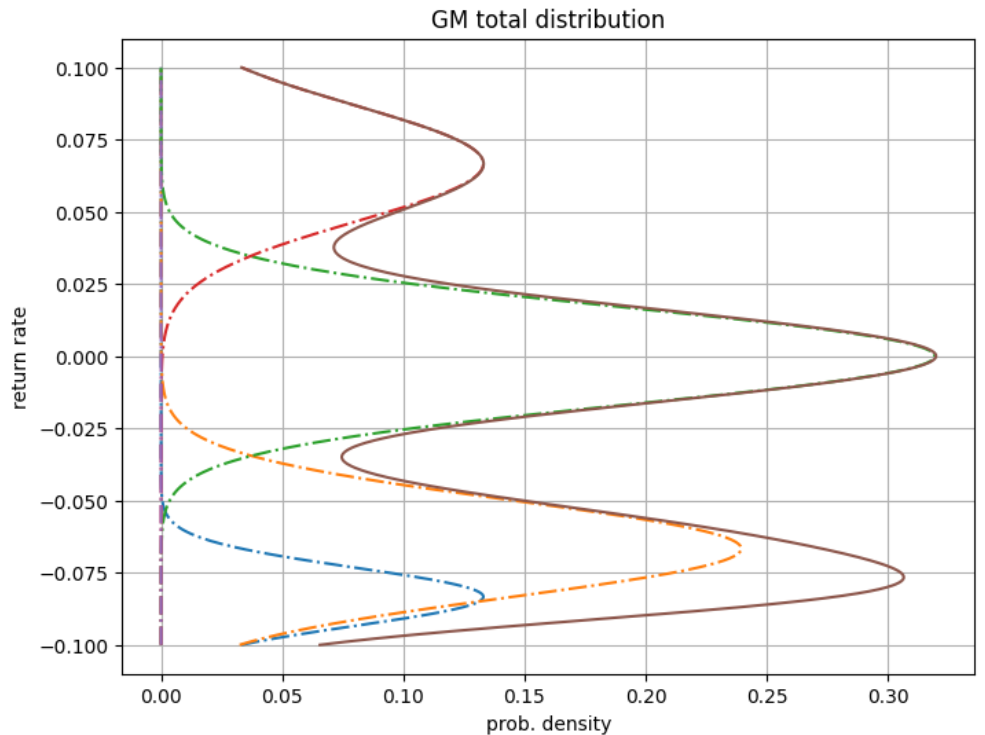}
        \caption{Multimodal distribution}
    \end{subfigure}
    % 改进后的注释
    \caption{Mixture of five Gaussian distributions. The x-axis indicates the probability density, indicating how frequently a particular return rate is expected. The y-axis represents the return rate, with values in the range of -0.1 to 0.1 (indicating return rates in the next period from -10\% to 10\%). Each subfigure demonstrates a different type of distribution formed by a mixture of Gaussian components: (a) Normal distribution: This  mixture of Gaussian approximates a normal distribution with a positive mean.(b) Skewed distribution: This distribution demonstrates skewness, formed by the combination of several Gaussian components.(c) Bimodal distribution: This illustrates a bimodal shape, indicating the presence of two distinct peaks. (d) Multimodal distribution: This distribution has multiple peaks, showcasing a more complex multimodal structure.}
    % \parbox{\textwidth}{
    %     \vspace{0.3cm}  % 这里增加垂直空间，调整值以满足需求
    %     \footnotesize
    %     \textit{Notes.} The x-axis indicates the probability density, indicating how frequently a particular return rate is expected. The y-axis represents the return rate, with values in the range of -0.1 to 0.1 (indicating return rates in the next period from -10\% to 10\%). Each subfigure demonstrates a different type of distribution formed by a mixture of Gaussian components:
    %     (a) Normal distribution: This  mixture of Gaussian approximates a normal distribution with a positive mean.(b) Skewed distribution: This distribution demonstrates skewness, formed by the combination of several Gaussian components.(c) Bimodal distribution: This illustrates a bimodal shape, indicating the presence of two distinct peaks. (d) Multimodal distribution: This distribution has multiple peaks, showcasing a more complex multimodal structure.
    % }
    \label{fig:grid_of_images}
\end{figure}

Consider a mixture of five one-dimensional Gaussian distributions as an example. In \Cref{fig:grid_of_images}, the dotted lines in different colors represent the individual Gaussian distributions, while the solid line shows the probability density function of the Gaussian mixture distribution formed by these components. It is evident that the weighted sum of these components, with specific weights, results in a uniquely defined Gaussian mixture distribution. By varying the weights, the mixture distribution can exhibit a range of non-traditional distribution patterns.

A heightened market volatility following the release of significant information often leads to non-traditional distribution patterns in terms of stock return. For instance, during the early 2020s, global markets experienced significant uncertainty surrounding the impact of the COVID-19 outbreak. Initially perceived as a localized issue, the rapid spread of the virus and its implications for global economies led to heightened uncertainty and volatility in stock markets worldwide. Sectors such as airlines, travel, and healthcare experienced significant fluctuations as investors grappled with divergent interpretations of information and its implications for future economic conditions.

The variance of the Gaussian mixture distribution \(\sigma^2_t\) serves as a key measure of the distribution's volatility, calculated based on the variance and mean of each component's Gaussian distribution. This can be computed using Equation \eqref{eq:gmm_var_time}:
\begin{equation}
    \label{eq:gmm_var_time}
    \sigma_t^2 = \sum_{i=1}^{n} w_{i,t} \cdot (\mu_{i,t}^2 + \sigma_{i,t}^2) - \left(\sum_{i=1}^{n} w_{i,t} \cdot \mu_{i,t}\right)^2.
\end{equation}
The calculation of the variance is derived from concepts involving the mean of a mixture distribution and second-order central moments. The mean of the mixture distribution \(\mu_t\) represents a weighted average of the means of each component and is obtained by integrating the product of the mixture distribution function \(f_t(x)\) and \(x\):
\begin{align}
    \label{eq:pro1}
    \mu_t &= \int_{-\infty}^\infty x f_t(x) dx \\
          &= \sum_{i=1}^n w_{i,t} \int_{-\infty}^\infty x f_{i,t}(x) dx \\
          &= \sum_{i=1}^n w_{i,t} \cdot \mu_{i,t}.
\end{align}
The second-order central moment of the mixture distribution, \(\mathrm{E}[X^2]\), represents the expectation of the square of the continuous random variable \(X\) and reflects the breadth of the distribution and the thickness of the tails. These moments can be obtained by integrating \(x^2 f_t(x)\):
\begin{align}
    \label{eq:pro2}
    \mathrm{E}[X^2] &= \int_{-\infty}^\infty x^2 f_t(x) dx \\
                    &= \sum_{i=1}^n w_{i,t} \int_{-\infty}^\infty x^2 f_{i,t}(x) dx \\
                    &= \sum_{i=1}^n w_{i,t} \cdot (\mu_{i,t}^2 + \sigma_{i,t}^2).
\end{align}
Finally, the variance \(\sigma_t^2\) of the Gaussian mixture distribution can be obtained by subtracting the square of the mean from the second-order central moments, according to the law of total variance. This calculation captures the volatility of the mixture distribution:
\begin{align}
    \label{eq:pro3}
    \sigma_t^2 &= \mathrm{E}[X^2] - \mu_t^2 \\
               &= \sum_{i=1}^{n} w_{i,t} \cdot (\mu_{i,t}^2 + \sigma_{i,t}^2) - \left(\sum_{i=1}^{n} w_{i,t} \cdot \mu_{i,t}\right)^2.
\end{align}

\subsection{Neural Network Structure}
\label{sec:network structure}
Deep learning technology is undergoing rapid development, with various novel network structures and solutions constantly emerging. Compared with traditional machine learning algorithms, neural networks constructed by deep learning exhibit superior learning performance for handling complex information. However, due to the inherent complexity and black-box nature of deep learning networks, their prediction results often lack interpretability, especially in tasks with significant uncertainty risks.
  
The mixture distribution network effectively addresses this issue by providing not only a direct prediction mean for input information but also the potential probability distribution function of the prediction value. \cite{zhang_short-term_2019} utilizes mixture Gaussian model on wind turbine power uncertainty analysis and forecasting. This probability distribution represents the conditional probability of possible forecast values, offering insights into the uncertainty associated with the predictions. When historical stock price and trading volume are utilized as input data, this conditional probability distribution reflects the probability distribution function of future returns based on the current historical trend of the stock.

The advantage of deep learning networks lies in their capacity to handle nonlinear information within data, and various network structure variants have emerged to accommodate different input data structures. Among these, Crossformer stands out as a deep neural network structure specifically designed for processing time series data \cite{zhang_crossformer_2023}. Derived from the transformer network architecture, Crossformer overcomes the limitations of traditional Transformer networks in handling time series data by incorporating DSW embedding and TSA layers to construct a hierarchical encoder-decoder. This design effectively captures temporal and spatial dimensions in the data, enhancing the extraction of temporal information.

\begin{figure}
    \centering
    \includegraphics[width=0.9\textwidth]{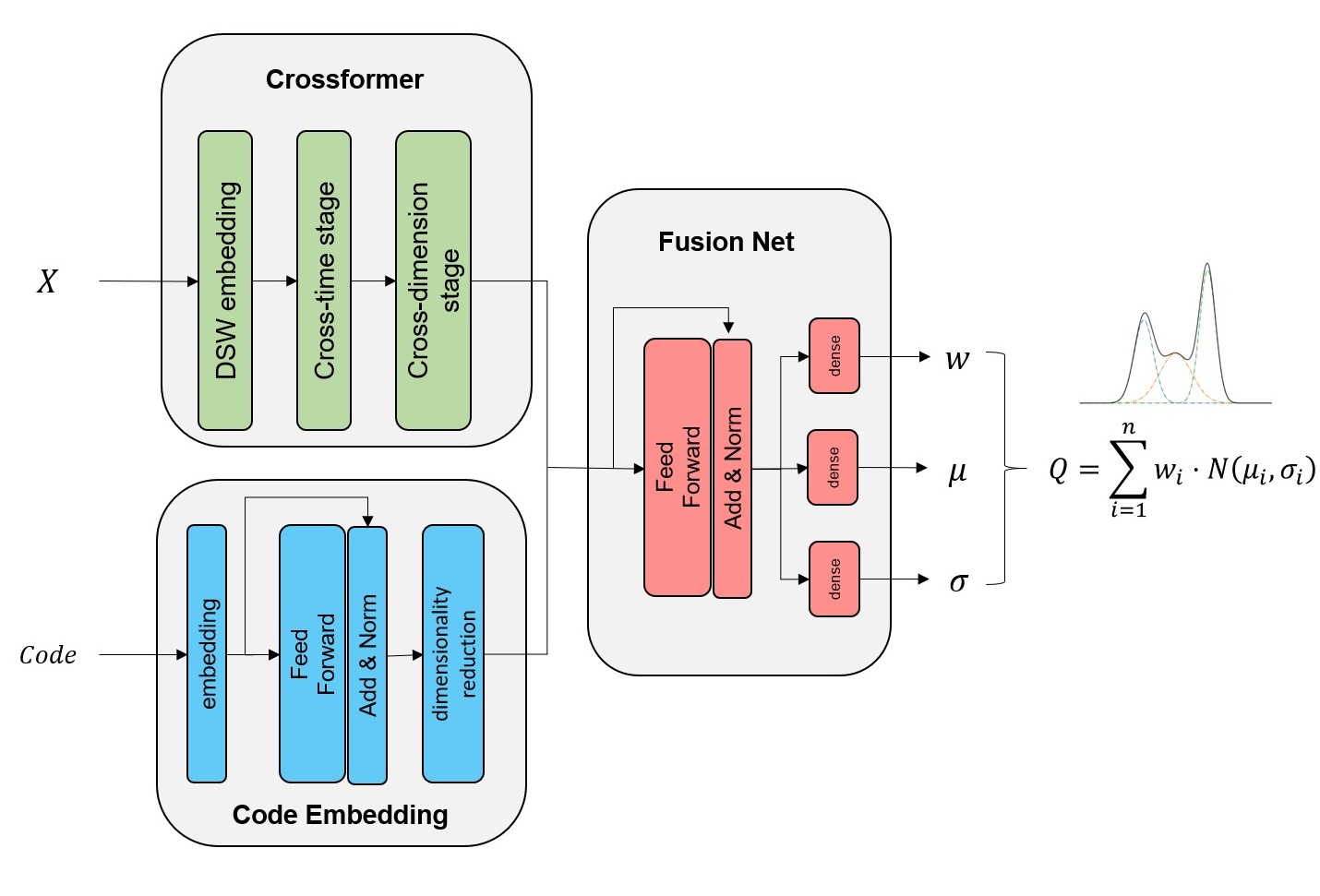}
    \vspace{-2ex}
    \caption{Deep learning neural network structure. This figure shows neural network structure. It consists of three main components: (a) Crossformer: This component processes indicators as matrix \(X \in \mathbb{R}^{C \times T}\) through three types of layers in Crossformer. (b) Code Embedding: This part handles input stock code by embedding, feed-forward network, add \& norm layer, and dimensionality reduction (dense layer). (c) Fusion Net: This module integrates the outputs from Crossformer and Code Embedding, passing through feed-forward network and add \& norm layers, producing final outputs $w$, $\mu$, and $\sigma$. The equation $Q = \sum_{i=1}^{n} w_i \cdot \mathcal{N}(\mu_i, \sigma_i)$ represents the probability function corresponding to the output parameters.}
    \label{fig:net_structure}
\end{figure}

Specifically, we adopt the mixture density network proposed by \cite{bishop_mixture_1994} to predict the probability distribution of asset returns, integrating it with the Crossformer network and deep embedding technology. As shown in \Cref{fig:net_structure}, we utilize ten technical indicator vectors \(X \in \mathbb{R}^{C \times T}\) as input data for the Crossformer network, where \(C\) represents the number of indicators and \(T\) represents the length of the historical time window. 

Simultaneously, the stock code be put into the code embedding network. In this process, we first map each unique stock code to a unique integer identifier, creating a vocabulary where each stock code corresponds to a specific integer index. Next, we build a vocabulary that contains all unique stock codes. We then initialize an embedding layer, which maps these integer indices to continuous embedding vectors. Finally, the integer sequence representing the stock codes is passed through the embedding layer, generating continuous embedding vectors for each stock code. This transformation allows the model to effectively capture and utilize the unique characteristics associated with each stock code during the prediction process.

Subsequently, the outputs of both networks are combined as input for the fusion network. The fusion network merges information. This information is from the technical indicators processed by the previous network and the corresponding stock code information. Through the maximum likelihood estimation method, we maximize the probability of the actual return value $y$ in the predicted probability density function $Q$, which takes $x$ as input data. We then use the negative log-likelihood as the loss function $\mathcal{L}$ to train the model parameters $\theta$. The objective function of the optimization is formulated as follows:

\begin{equation} \label{eq:obj}
    \min_{\theta} \mathcal{L} = \min_{\theta} -\sum_{k=1}^{m} \log \left(Q(y|x_k;\theta)\right)
\end{equation}
subject to
\begin{equation}
\quad w_i(x; \theta) \geq 0,\quad\sigma_i(x; \theta) > 0,\quad\sum_{i=1}^{n} w_i(x; \theta) = 1, 
\end{equation}
where
\begin{equation}
    Q(y|x; \theta) = \sum_{i=1}^{n} w_i(x; \theta) \mathcal{N}(y|\mu_i(x; \theta), \sigma_i(x; \theta)), 
\end{equation}

\begin{equation}
    \mathcal{N}(y|\mu_i, \sigma_i) = \frac{1}{\sqrt{2 \pi \sigma_i^2}} \exp \left( -\frac{(y - \mu_i)^2}{2 \sigma_i^2} \right).
\end{equation}
In eq.~\eqref{eq:obj}, $m$ represents the total number of samples in the optimization process, probability function $Q$ is influenced by input samples data $x$, and $\theta$ is a set of model parameters. This procedure outputs the optimized hyper-parameters of the Gaussian mixture distribution: the mean vector $\mu$, standard deviation vector $\sigma$, and weight vector $w$. These parameters are essential for constructing the Gaussian mixture distribution, determining the shape of the probability density function. The length of the vectors is determined by the number $n$ of Gaussian distribution components in Gaussian mixture distribution, allowing the mixture distribution to fit a wide range of distribution shape. In the experiment part, we will adopt $n=9$, a mixed distribution composed of nine Gaussian distributions for the experiment.

\subsection{Code Embedding and t-SNE Visualization}
\label{sec:code embedding and t-sne visualization}

\begin{figure}
    \centering
    \includegraphics[width=1.0\textwidth]{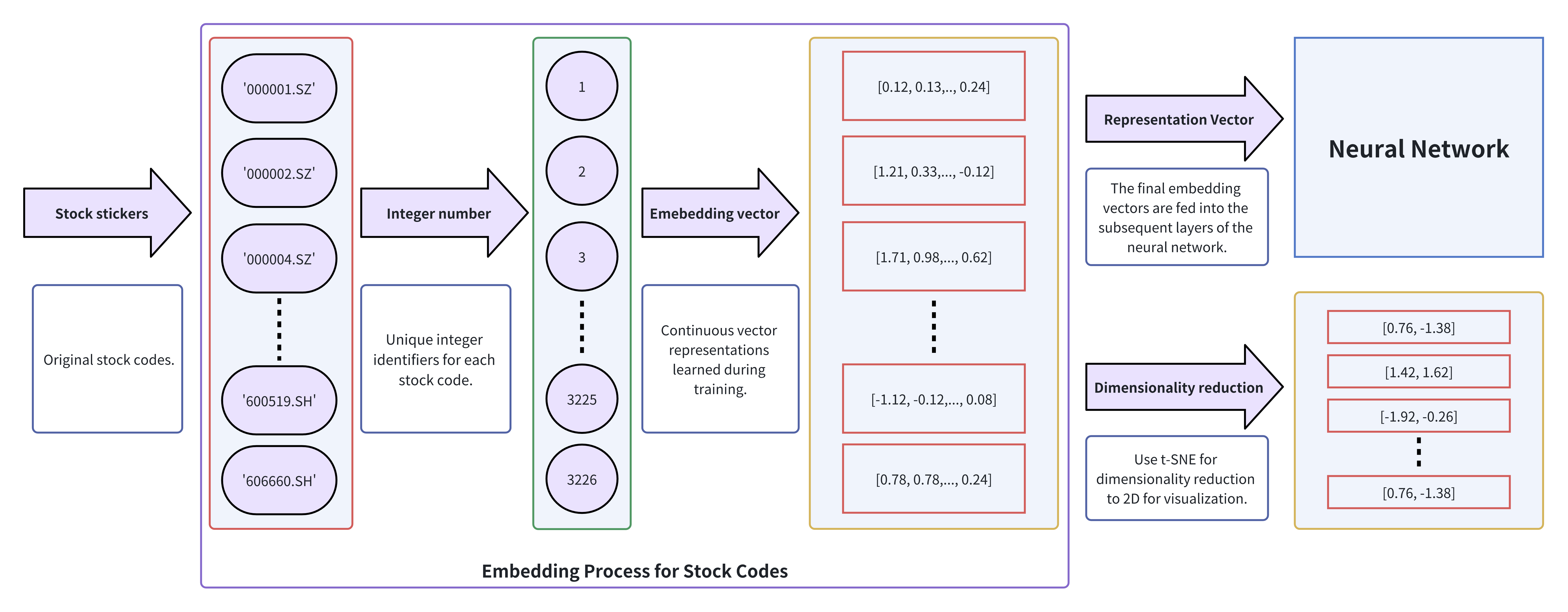}
    \caption{Stock code embedding workflow. This figure shows the process of embedding stock codes into continuous vectors. This process uses the Bag-of-Words approach for stock codes, where each unique stock code is treated as a distinct token, mapped to an integer, and then embedded into a continuous vector space. The embedding vectors are then fed into subsequent layers of the neural network or visualized by dimensionality reduction techniques.}
    \label{fig:embedding_process}
\end{figure}

The input samples of the neural network consist of thousands of stocks. To help the model learn the distinctive properties of each stock, we jointly embed the stock ticker vector \(Code\) corresponding to the sample indicator \(X\) into the network. The workflow shown in \Cref{fig:embedding_process} illustrates how stock codes are embedded into continuous vectors. This method applies the Bag-of-Words model to stock codes, treating each unique stock code as a distinct word. Each stock code is mapped to an integer and then embedded into a continuous vector space. These vectors are learned during training and capture the unique features of each stock code. The learned vectors are then passed through the subsequent layers of the neural network for further processing.
Additionally, dimension reduction techniques, such as t-SNE, are used to map the high-dimensional embedding vectors to two dimensions for visualization, providing an intuitive understanding of stock code relationships and clusters. Previous studies \citep{mikolov_efficient_2013, levy2015improving, llinares2023deep, SCHWENK2007492} have shown that vector representations through deep learning can be applied to various tasks, such as measuring syntactic and semantic word similarities and protein sequence alignment. Similarly, in our neural network training process, stock tickers with similar risk attributes tend to be located closer in the vector space, while tickers with significant attribute differences are positioned farther apart. This process enables the model to learn and capture the relative relationships and dynamic patterns among stocks based on their technical indicators.

% To visualize the distances among the samples on a certain feature in the vector space, \cite{maaten_visualizing_2008} proposed the t-SNE technique. This technique reduces the dimensionality of samples from high-dimensional space to two-dimensional space for visualization while preserving relative distances among those samples. We use t-SNE on the vector representations of stock codes. This process helps us to clearly see which stocks have similar risk attributes by looking at their positions in the two-dimensional plot. After mapping all stock tickers into a two-dimensional space via t-SNE dimensionality reduction, we draw a 2-D scatter plot, which employs the two dimensions of the vector space as the x-axis and y-axis of the scatter plot. To further analyze similarity among stocks, we use graduated colors to label the magnitude of several risk-related attribute values of each stock in the past two years. The clustering of colors intuitively display the similarity or difference between stocks based on these attributes. This method provides the neural network with the risk characteristics at the individual stock level, and at the same time provides a new perspective for observing the risk similarity among assets.

To visualize the distances between samples based on certain features in the vector space, \cite{maaten_visualizing_2008} introduced the t-SNE technique. This method reduces the dimensionality of high-dimensional data to two dimensions, preserving the relative distances between samples for visualization. We apply t-SNE to the vector representations of stock codes, allowing us to clearly identify stocks with similar risk attributes by observing their positions in the two-dimensional plot. After mapping all stock tickers into a two-dimensional space using t-SNE, we create a scatter plot, with the two dimensions of the vector space serving as the x- and y-axes. To further analyze stock similarity, we apply graduated colors to represent the magnitude of various risk-related attributes for each stock over the past two years. The clustering of colors visually indicates the similarities or differences between stocks based on these attributes. This method not only provides the neural network with risk characteristics at the individual stock level but also offers a new perspective on risk similarities among assets.

\subsection{Evaluation Metrics}
\label{sec:evaluation metrics}

Given that the time-varying distribution of price changes and the actual volatility cannot be observed directly, we employ the realized range volatility (RRV) metric as an approximate proxy for actual volatility. This method is commonly used in risk assessment. This proxy is then utilized to assess the similarity between the model-predicted volatility and the substitute of actual volatility using either the mean square error (MSE) or the quasi-likelihood (QLIKE) evaluation criterion as described in Equations \eqref{eq:mse} and \eqref{eq:qlike} respectively,
\begin{equation}
    \label{eq:mse}
    \mathrm{MSE} = \frac{1}{m} \sum_{i=1}^{m} (\sigma_i - \hat{\sigma}_i)^2,
\end{equation}
\begin{equation}
    \label{eq:qlike}
    \mathrm{QLIKE} = \frac{1}{m} \sum_{i=1}^{m} \left( \frac{\sigma_i}{\hat{\sigma_i}} + \log(\hat{\sigma_i}) \right),
\end{equation}
where $m$ represents the number of samples, $\sigma_i$ represents the actual volatility of the i-th sample (proxied by RRV), $\hat{\sigma_i}$ represents the corresponding predicted volatility. Despite both being disparity indicators between the model predicted and the realized volatilities,the QLIKE metric, different from the MSE metric, pays particular attention to the deviation of the ratio of the predicted to the true value, imposing substantial penalties for under- or over-predicted volatilities, thus particularly useful in evaluating the accuracy of volatility prediction models, especially regarding their sensitivity to extreme values. In practice, smaller values of the MSE and QLIKE metrics indicate smaller discrepancies between forecast and actual volatility, whereas larger values imply greater differences.

Approximating actual volatility with RRV may encounter large deviations under certain circumstances. For instance, volatility estimates may be influenced by the choice of measurement windows, with short-term windows possibly failing to capture long-term trends, and long-term windows potentially obscuring short-term fluctuations. Moreover, the common assumption that log returns follow a normal distribution may not always align with market realities, especially during periods of dramatic short-term movements in the market or specific stocks~\citep{ANDERSEN200143}. These extreme events can cause actual volatility to diverge significantly from historical levels over brief periods, and the RRV calculations based on historical data may not promptly capture these volatility changes. Besides, for the actual distribution of returns exhibiting skewed or heavy-tailed characteristics, the RRV calculations may inadequately reflect the market condition.

To comprehensively evaluate a model's effectiveness in predicting market volatility, we also employ the continuous ranked probability score (CRPS) metric~\citep{gneiting2007crps}. The CRPS enables a comprehensive assessment of the entire forecast distribution by comparing the agreement between the predictive distribution's probability density function and the actual occurrence values, in terms of central tendency and distribution width. A smaller CRPS value indicates closer alignment between the model's return predicted distribution and the actual occurrence, reflecting the overall quality and reliability of the model predictions. The mathematical formula for the CRPS indicator is given by Equation \eqref{eq:crps},
\begin{equation}
    \label{eq:crps}
    \operatorname{CRPS}=\int_{-\infty}^{\infty}(F(y)-\mathbf{1}_{y \geq x})^2 \mathrm{~d}y,
\end{equation}
where $F$ represents the cumulative distribution function of the predictive Gaussian mixture distribution, indicating the probability that the random variable is less than or equal to a specific value $y$. The variable $y$ spans all potential outcomes, while $x$ represents the actual return. The indicator function $\textbf{1}_{y \geq x}$ evaluates to 1 when $y$ is greater than or equal to the observed return $x$, and 0 otherwise.

The CRPS metric assesses the overall accuracy of the predictive distribution by computing the square of the difference between the predictive cumulative distribution function $F(y)$ and the indicator function at the actual observation $x$, and then integrating this squared difference over all possible values. This integration yields a non-negative value, with values closer to 0 indicating greater proximity between the predictive distribution and the actual observations. Consequently, the CRPS does not need to use any proxies of the actual volatility such as the RRV indicator. In practice, due to the complexity of the mathematical formulation of the Gaussian mixture distribution, we use Monte Carlo simulation to compute its CRPS value. Specifically, we perform 2000 samples for each Gaussian mixture distribution function and calculate the CRPS value for the sampled points.

\section{Empirical Results}
\label{sec:empirical results}

\subsection{Evaluation Setup}
In this study, we introduce a mixture density network model designed specifically for processing time series data, termed as MDNe, which enhances prediction accuracy by incorporating code embedding technique. To visualize the impact of code embedding on performance, we also examine a control model, denoted as MDN, which does not utilize code embedding. This comparative analysis enables us to evaluate the effectiveness of the code embedding technique.

Furthermore, we compare these two models with the traditional family of GARCH models, including GARCH, GJR-GARCH, APARCH, and TARCH models. To ensure fair comparisons, we simplify all GARCH models to univariate forms and set the parameters p and q uniformly to 1. This standardization enables us to assess each model's performance consistently under identical conditions, with a focus on their ability to capture volatility rather than model complexity. We adopt a zero-mean model for all stock return series to prioritize the analysis of GARCH model volatility, avoiding potential distractions from mean changes and focusing on volatility prediction capabilities. The zero-mean model for GARCH and its variants does not affect its volatility measurement, thus will not influence the calculation of the MSE and QLIKE indicators.

In the evaluation phase, we assess the probability distribution function of asset returns obtained from all model predictions using MSE, QLIKE, and CRPS metrics. Specifically, MSE and QLIKE gauge the discrepancy between model-predicted volatility and the RRV, while CRPS metrics evaluate the consistency between model-predicted distributions and actual return values, ensuring assessments align with actual volatility.

Model training and hyperparameter optimization are performed using data from 2018 to 2020, with validation data spanning early 2021 to late 2022 for predicting volatility and potential return distributions for the following day. The dataset comprises historical price and volume data of 3,226 stocks, encompassing companies listed on China's A-share main board market prior to 2020, ensuring broad data coverage and robust findings.

\begin{figure}
    \centering
    \begin{subfigure}[b]{0.45\linewidth}
        \captionsetup{font=small}  % 局部设置字体大小
        \includegraphics[width=\linewidth]{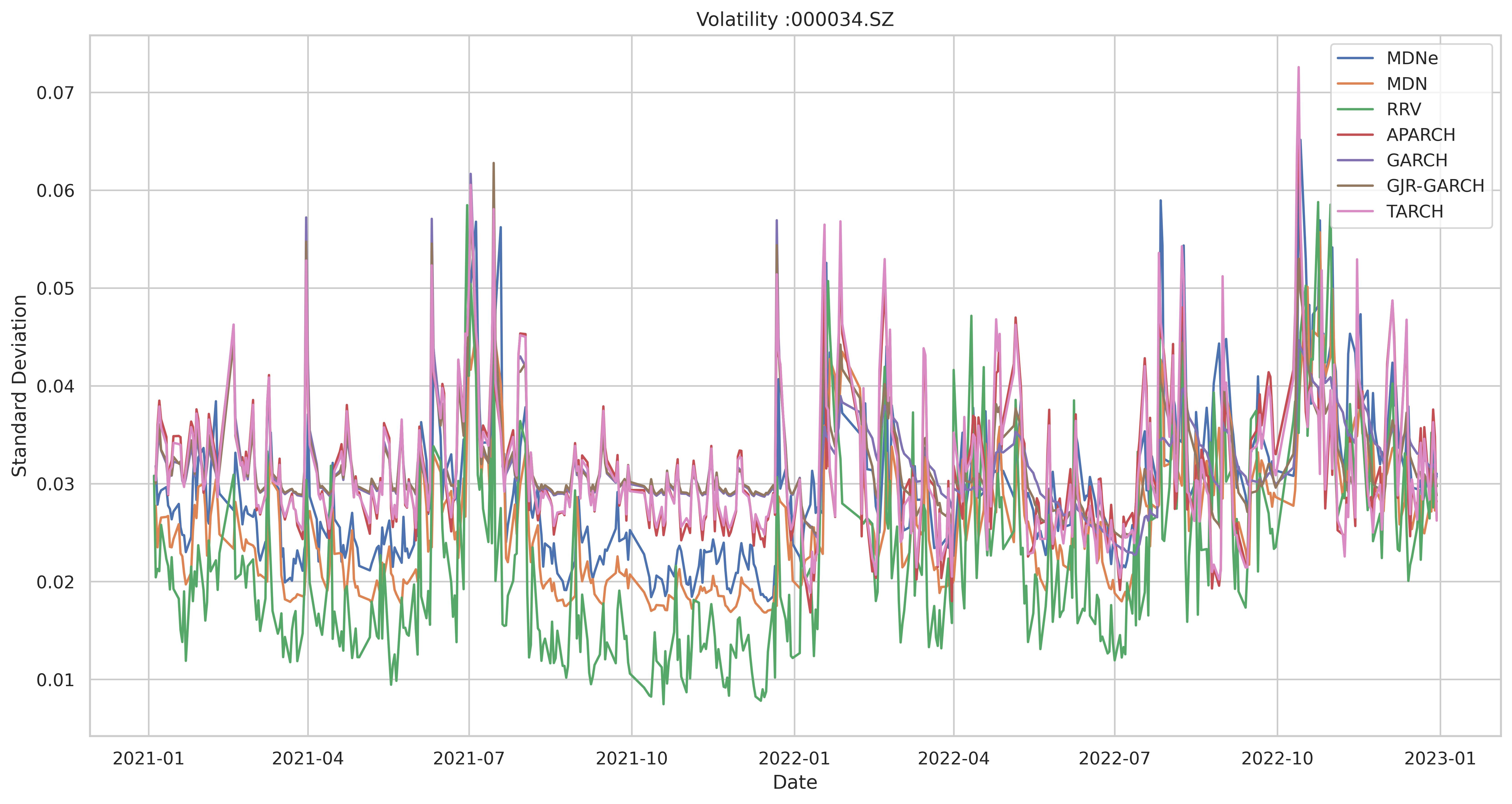}
        \caption{stock code:000034.SZ}
    \end{subfigure}
    \begin{subfigure}[b]{0.45\linewidth}
        \captionsetup{font=small}  % 局部设置字体大小
        \includegraphics[width=\linewidth]{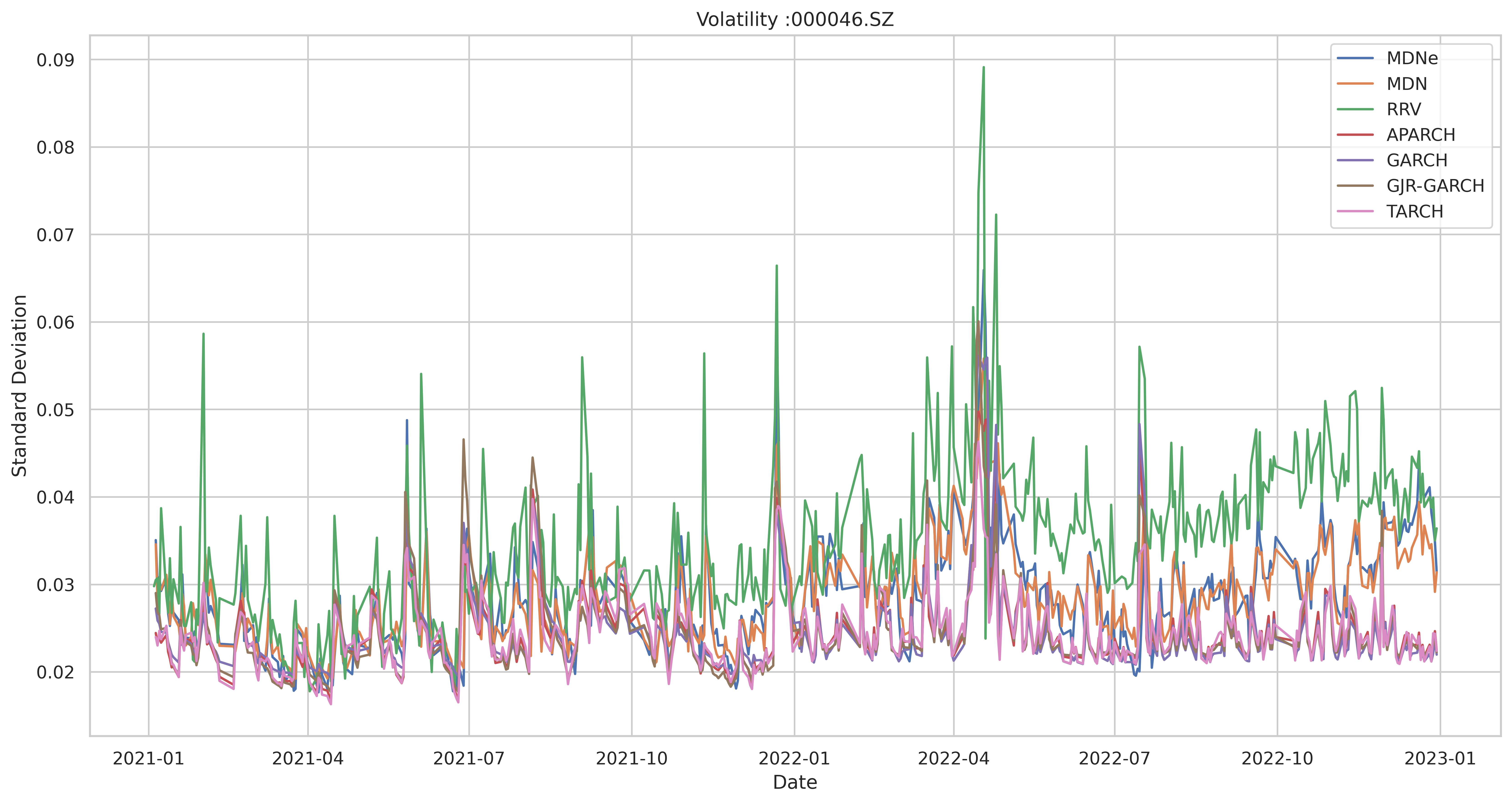}
        \caption{stock code:000046.SZ}
    \end{subfigure}
    
    \begin{subfigure}[b]{0.45\linewidth}
        \captionsetup{font=small}  % 局部设置字体大小
        \includegraphics[width=\linewidth]{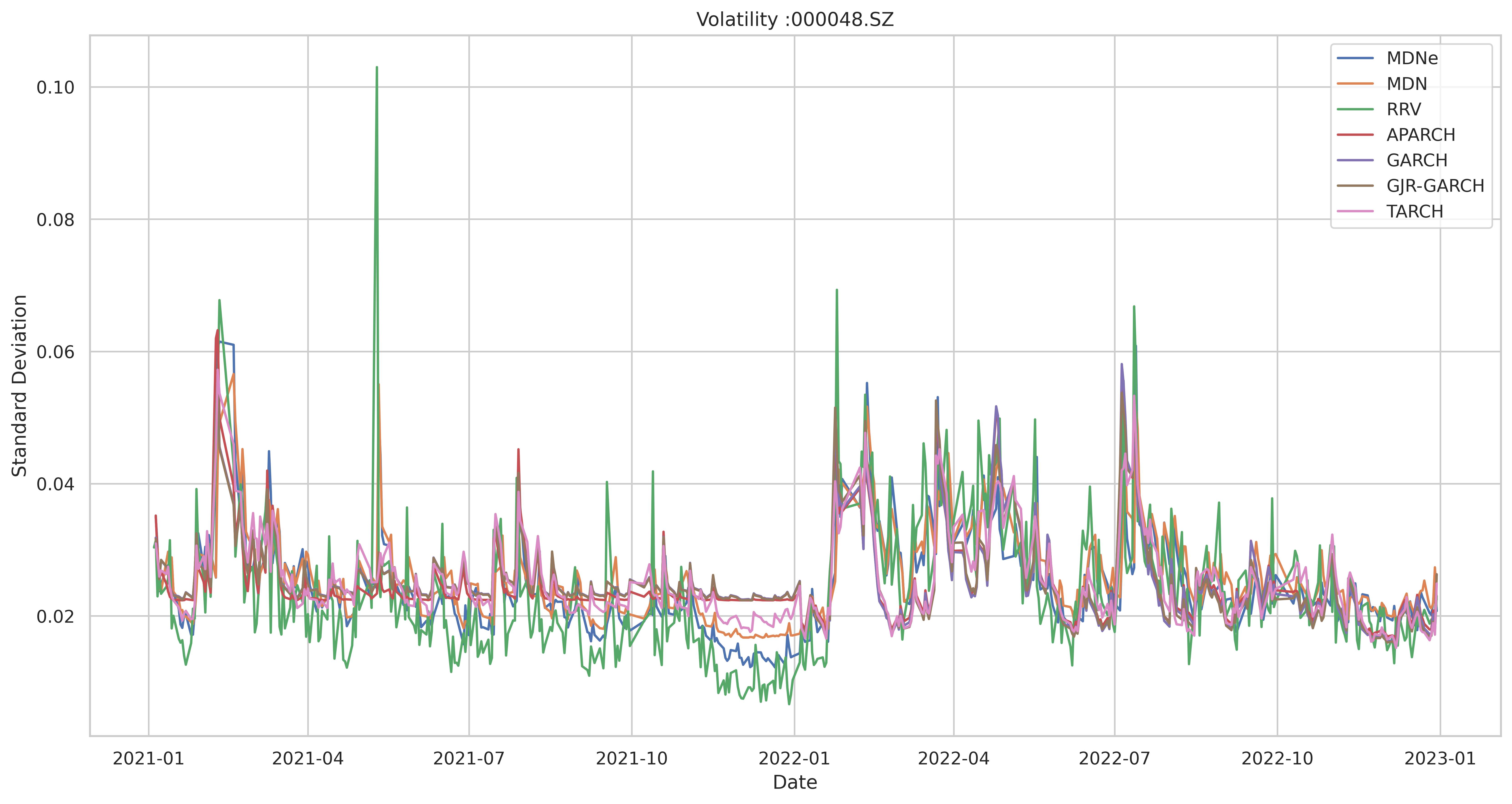}
        \caption{stock code:000048.SZ}
    \end{subfigure}
    \begin{subfigure}[b]{0.45\linewidth}
        \captionsetup{font=small}  % 局部设置字体大小
        \includegraphics[width=\linewidth]{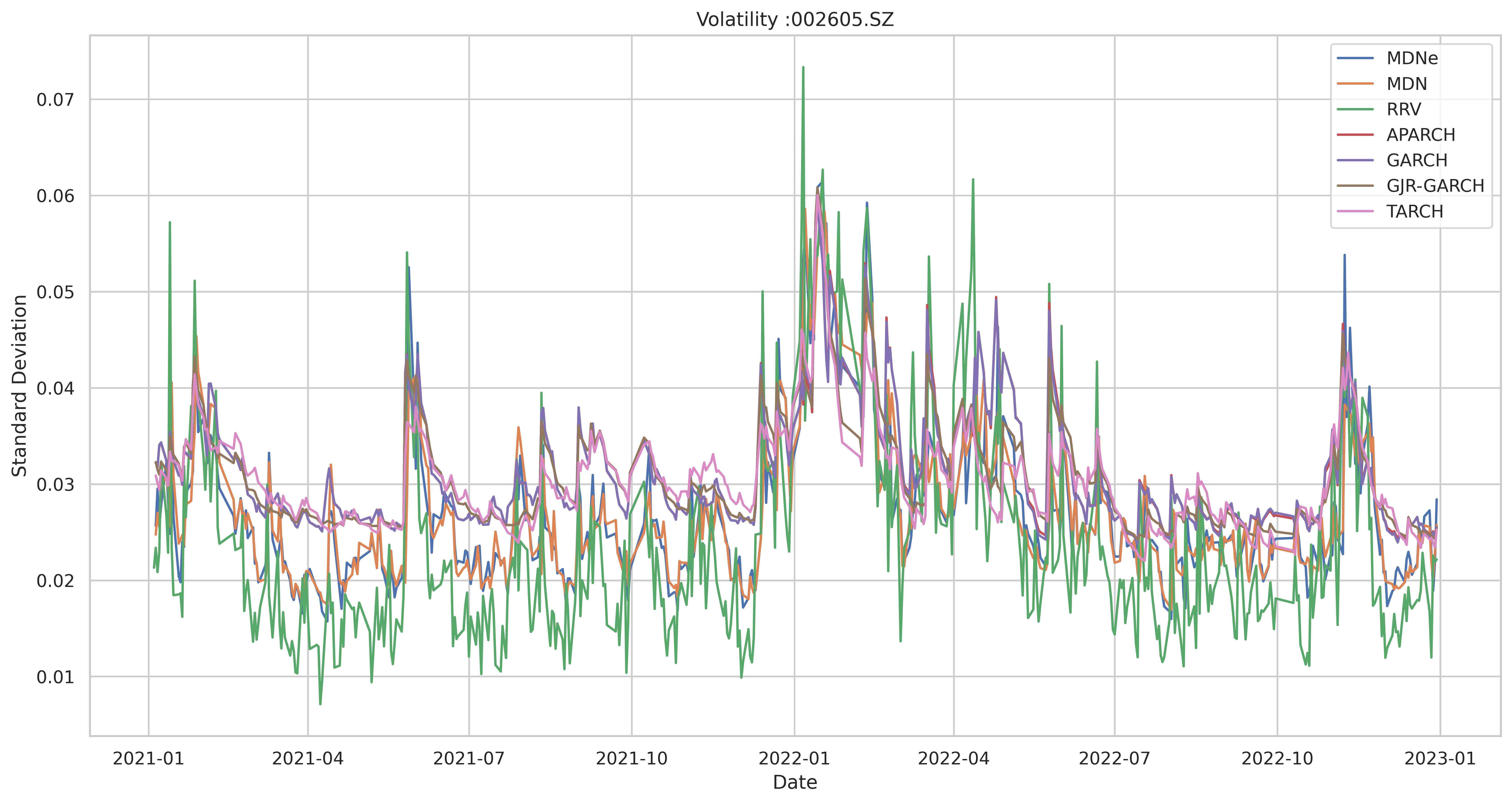}
        \caption{stock code:002605.SZ}
    \end{subfigure}
    \caption{Comparison of standard deviation for partial stocks. The figure compares the standard deviation for four stock samples, showing the performance of different volatility forecasting models. The green line represents the RRV indicator, serving as a proxy for actual volatility. The blue and yellow lines represent the volatility forecasts from the MDNe and MDN models, respectively. The other lines show forecasts from the GARCH model and its variants. It is observed that the MDN and MDNe models' forecasts are closer to the RRV curve.}
    \label{fig:val sample}
\end{figure}

We employ the annual rolling training, but there are differences in the training methods for the mixture density network and the GARCH-type models. For MDN and MDNe models, leveraging the deep learning framework's scalability and generalization capabilities, we aggregate data from all 3,226 stocks and train a single network model on the combined dataset, enabling the model to capture volatility patterns across stocks. For the GARCH family of models, we construct separate models for each stock, given their design for analyzing the volatility characteristics of individual time series. While this approach offers advantages in model fine-tuning, it becomes cumbersome and limits generalization across stocks when dealing with a large number of assets. We present volatility forecast curves for selected stocks in \Cref{fig:val sample}.

In the \Cref{fig:val sample}, as labeled, the green line represents the time series curve of the RRV indicator, which serves as a proxy for actual volatility. The blue and yellow lines respectively represent the time series curves of volatility forecasts obtained from the MDNe and MDN models, while the other four curves represent the time series curves of volatility forecasts from GARCH and its variants. It can be observed that the volatility curves obtained from the MDN and MDNe models are closer to the RRV curve. Although the volatility curve predicted by the GARCH model aligns with the RRV curve at an overall level, in terms of short-term fluctuations depicted in the RRV curve, the volatility curves from the MDN and MDNe models are more consistent with it. On the other hand, the curves from the GARCH model exhibit greater deviations from the RRV curve. While we have only selected samples of four stocks due to space limitations, these findings can be observed in most stock samples. Although the RRV curve cannot serve as a perfect substitute for actual volatility, it provides a basis for intuitively comparing volatility forecasts. Next, we will further evaluate the overall samples in detail using multiple evaluation metrics.

\subsection{Evaluation Results}
\label{sec:Model Evaluation}
In the analysis of the prediction performance of the six models presented in \Cref{tb:metrics}, several noteworthy observations emerge. Firstly, the MDNe model demonstrates superior predictive accuracy overall, particularly evident in the three core evaluation metrics: CRPS, MSE, and QLIKE, all of which yield the lowest predictive loss values. This robust performance underscores the significant advantage of the MDNe model in terms of comprehensive measures of forecast accuracy, emphasizing the efficacy of deep learning in volatility forecasting within financial markets.

\begin{table}
    \centering
    \small
    \caption{Evaluation of out-of-sample volatility forecasts.}
    \label{tb:metrics} 
    \begin{tabularx}{\textwidth}{lXXXXXX}
    \toprule
    Forecasting horizon & MDNe & MDN & GARCH & GJR-GARCH & APARCH & TARCH \\
    \midrule
    \underline{Total stock average:} & & & & & & \\
    CRPS (\%) & \textbf{1.4683} & 1.4701 & 1.4856 & 1.4862 & 1.4881 & 1.4865 \\
    MSE & \textbf{0.7233} & 0.7526 & 0.8942 & 0.8539 & 1.1537 & 0.8896 \\
    QLIKE & \textbf{1.8501} & 1.8553 & 1.8657 & 1.8661 & 1.8664 & 1.8650 \\
    
    \underline{10\% quantile}: & & & & & & \\
    CRPS (\%) & 1.0358 & \textbf{1.0335} & 1.0502 & 1.0518 & 1.0522 & 1.0518 \\
    MSE & 0.3542 & 0.3825 & 0.3597 & 0.3789 & \textbf{0.3528} & 0.3607 \\
    QLIKE & \textbf{1.5976} & 1.6072 & 1.6070 & 1.6089 & 1.6054 & 1.6064 \\
    
    \underline{20\% quantile:} & & & & & & \\
    CRPS (\%) & 1.1786 & \textbf{1.1733} & 1.2002 & 1.1999 & 1.1988 & 1.1989 \\
    MSE & \textbf{0.4425} & 0.4515 & 0.4710 & 0.4848 & 0.4675 & 0.4668 \\
    QLIKE & \textbf{1.7039} & 1.7077 & 1.7111 & 1.7119 & 1.7111 & 1.7117 \\
    
    \underline{50\% quantile:} & & & & & & \\
    CRPS (\%) & 1.4675 & \textbf{1.4659} & 1.4819 & 1.4824 & 1.4835 & 1.4831 \\
    MSE & 0.6512 & \textbf{0.6474} & 0.7559 & 0.7620 & 0.7568 & 0.7292 \\
    QLIKE & 1.8685 & \textbf{1.8683} & 1.8803 & 1.8802 & 1.8791 & 1.8789 \\
    
    \underline{80\% quantile:} & & & & & & \\
    CRPS (\%) & \textbf{1.7652} & 1.7722 & 1.7878 & 1.7855 & 1.7877 & 1.7858 \\
    MSE & \textbf{0.9645} & 0.9703 & 1.1662 & 1.1856 & 1.1744 & 1.1396 \\
    QLIKE & 2.0329 & \textbf{2.0325} & 2.0384 & 2.0396 & 2.0403 & 2.0383 \\
    
    \underline{90\% quantile:} & & & & & & \\
    CRPS (\%) & \textbf{1.9406} & 1.9489 & 1.9556 & 1.9563 & 1.9565 & 1.9566 \\
    MSE & \textbf{1.1790} & 1.2288 & 1.4662 & 1.4816 & 1.4860 & 1.4589 \\
    QLIKE & 2.1154 & \textbf{2.1125} & 2.1256 & 2.1263 & 2.1295 & 2.1257 \\
    \bottomrule
    \end{tabularx}
    \parbox{\textwidth}{
        \vspace{0.3cm}  % 这里增加垂直空间，调整值以满足需求
        \footnotesize
        \textit{Notes.} This table evaluates the out-of-sample volatility forecasting performance of six models: MDNe, MDN, GARCH, GJR-GARCH, APARCH, and TARCH. The table is structured to show the performance across different quantiles (10\%, 20\%, 50\%, 80\%, and 90\%) and includes three metrics: CRPS, MSE, and QLIKE. Each section within the table presents the average performance of the models for the total stocks and different volatility levels. The bold values indicate the best performance that model achieves the lowest predictive loss for the respective metric and quantile.}
\end{table}

By comparing the predictive performance of samples at different percentiles of the volatility level, it becomes apparent that the MDN model outperforms other models in low volatility quantile level, such as the 10th percentile, suggesting a subtle advantage in capturing lower levels of volatility.

Although the loss value of the evaluation indicators increases with the increase of the volatility level, it can be found that the loss values of MDN and MDNe are still smaller than the GARCH model and its variants. This indicates that the performance of the mixture distribution network is still work well under strong volatility levels. Besides, the CRPS and the MSE loss values of MDNe are smaller than those of MDN at high volatility levels, while the QLIKE loss value is almost equal, which means that MDN with code embedding technique can take advantage on using individual stock information to improve the forecasting accuracy.

The observed discrepancy in forecasting performance between GARCH models and deep learning-based MDNe and MDN models may be attributed to the inherent limitations of GARCH models, which rely on simplifying assumptions about volatility aggregation phenomena. In contrast, real-world financial markets exhibit more complex dynamics and nonlinear features during extreme events and periods of heightened volatility, which are challenging for GARCH models to accurately capture. The MDNe and MDN models leverage their advantage on nonlinear modeling to capture these complex market dynamics, particularly in the prediction of high volatility levels.

In order to evaluate the performance of these volatility prediction models more intuitively, we employed the Diebold-Mariano (DM) test~\citep{Francis1995} to compare the differences in the predictive performance between models. This statistical method assesses whether there are significant differences in prediction accuracy between two models by comparing their prediction error sequences. The results of the DM-test are presented in \Cref{tb:dmtest}. If the t-statistic value is positive, it indicates that the model corresponding to the column of that value outperforms the model corresponding to the row of that value in terms of prediction performance. Conversely, if the t-statistic value is negative, it indicates that the model corresponding to the row performs better than the model corresponding to the column. Additionally, if the absolute value of the t-statistic exceeds 2.58, it indicates a statistically significant difference between the two models at a significance level of 1\%. A higher t-statistic value implies a more significant advantage of one model over the other. In our analysis, we mainly focus on evaluating the statistical significance differences in model predictions using MSE and QLIKE as loss functions.

\begin{table}
\centering
\small
\caption{Diebold-Mariano test results.}
\label{tb:dmtest}
\begin{tabularx}{\textwidth}{XXXXXX}
\toprule
Forecasting horizon & MDNe & MDN & GARCH & GJR-GARCH & APARCH \\
\midrule
\underline{MSE Loss} & & & & & \\
MDN & 12.44** & & & & \\
GARCH & 84.29** & 72.8** & & & \\
GJR-GARCH & 87.84** & 75.62** & 22.82** & & \\
APARCH & 21.83** & 21.16** & 15.50** & 14.78** & \\
TARCH & 42.02** & 38.65** & 12.62** & 9.04** & -14.41** \\
\midrule
\underline{QLIKE Loss} & & & & & \\
MDN & 42.23** & & & & \\
GARCH & 134.88** & 109.19** & & & \\
GJR-GARCH & 138.59** & 113.08** & 20.69** & & \\
APARCH & 129.76** & 107.43** & 17.26** & 6.17** & \\
TARCH & 120.80** & 96.08** & -17.38** & -32.82** & -28.99** \\
\bottomrule
\end{tabularx}
\parbox{\textwidth}{
    \vspace{0.3cm}  % 这里增加垂直空间，调整值以满足需求
    \footnotesize
    \textit{Notes.} This table presents the DM-test results comparing the performance of volatility forecasting models using MSE and QLIKE loss metrics. A positive t-statistic value indicates that the model in the corresponding column performs better than the model in the corresponding row, and vice versa. Standard errors are denoted by * \( p<0.05 \),** \( p<0.01 \).}
\end{table}

The \Cref{tb:dmtest} indicates that the DM-test results for MSE and QLIKE as loss metrics are consistent. Among them, MDNe has a significant advantage over other models. Specifically, when comparing MDNe with the MDN model, the t-statistic value for MSE loss is a positive 12.44, indicating that at the 1\% significance level, the MDNe model is better than the MDN model. At the same time, when comparing MDNe with the GARCH model and other variants, it also shows a large positive t-statistic value, indicating that there is also a significant advantage for this model. 

Moreover, when comparing the two mixture distribution network models, MDNe and MDN, with the GARCH model and its variants, the t-statistic values turn out to be very large. These t-statistic values are much greater than the t-statistic values that comparing between the GARCH-type models.This indicates that the prediction advantage of the mixture distribution network models over the GARCH models is larger than the difference in prediction advantage among the GARCH models. Nonetheless, the TARCH model performed better than other GARCH models in the DM-test results.

\subsection{Similarity of Return Uncertainty}
\label{sec:Similarity of return uncertainty}

Understanding the similarity in uncertainty surrounding stock returns is pivotal for effective portfolio management and risk mitigation strategies. By integrating deep embedding technique with t-SNE dimensionality reduction visualization, we gain a potent tool to uncover underlying market dynamics. These techniques shed light on the intrinsic relationships among stocks, offering novel insights for refining investment approaches and minimizing risks.

\begin{figure}
    \centering
    \begin{subfigure}[c]{0.32\linewidth}
        \captionsetup{font=small}  % 局部设置字体大小
        \includegraphics[width=\linewidth]{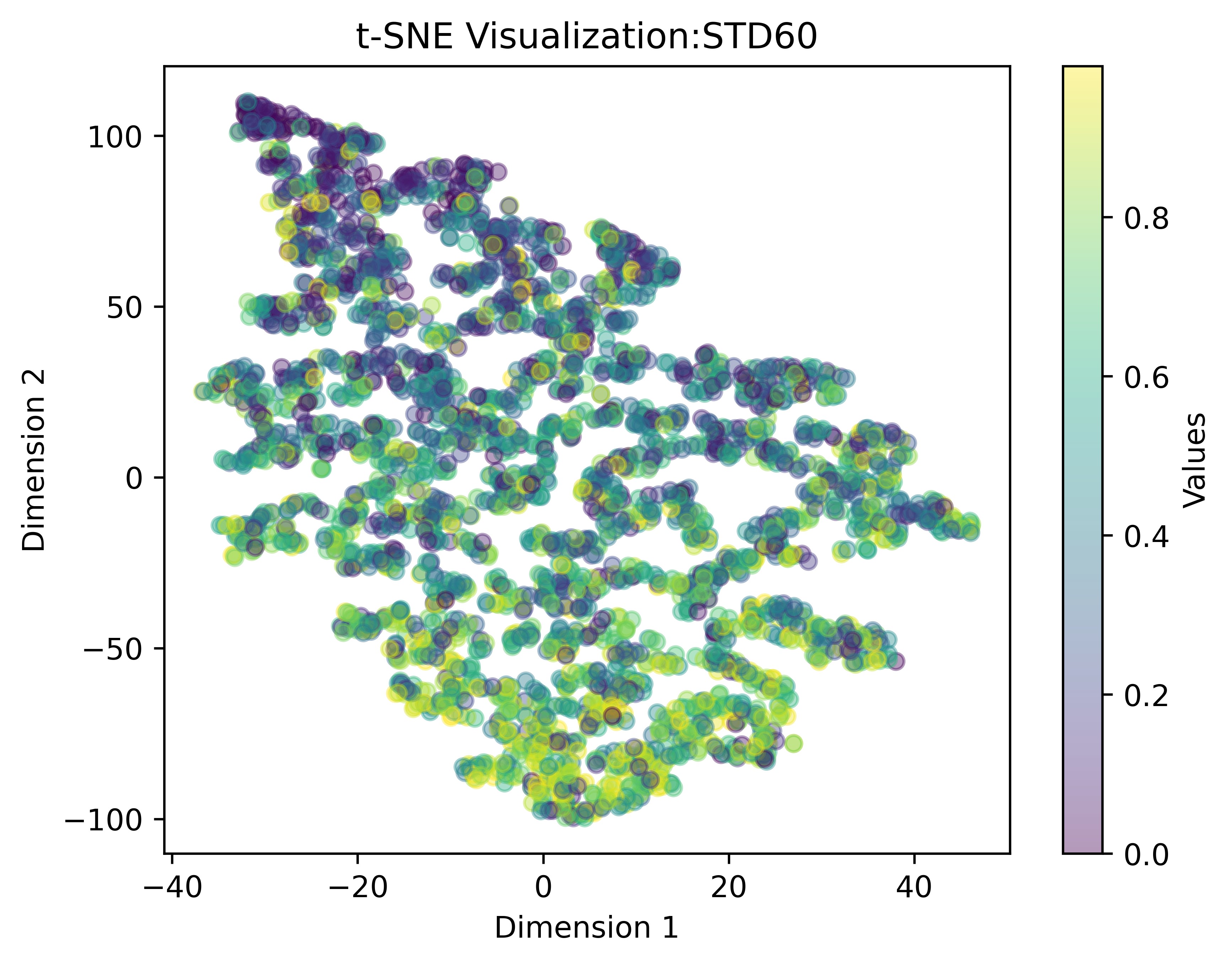}
        \caption{STD60.}
        \label{fig:sub1}
    \end{subfigure}
    % \hfill  % 使用hfill来在两个子图之间插入空格
    \begin{subfigure}[c]{0.32\linewidth}
        \captionsetup{font=small}  % 局部设置字体大小
        \includegraphics[width=\linewidth]{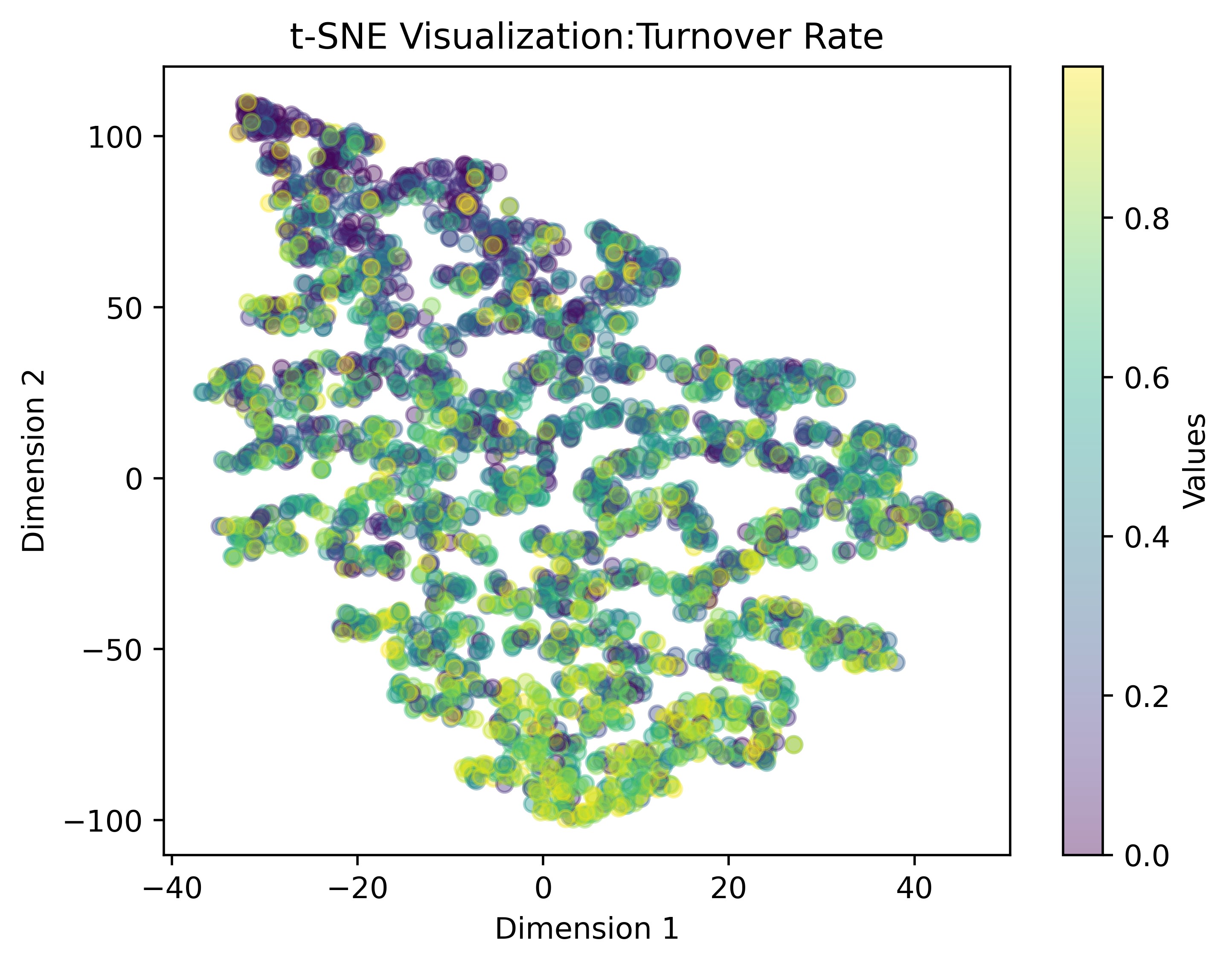}
        \caption{Turnover rate.}
        \label{fig:sub2}
    \end{subfigure}
    % \hfill
    \begin{subfigure}[c]{0.32\linewidth}
        \captionsetup{font=small}  % 局部设置字体大小
        \includegraphics[width=\linewidth]{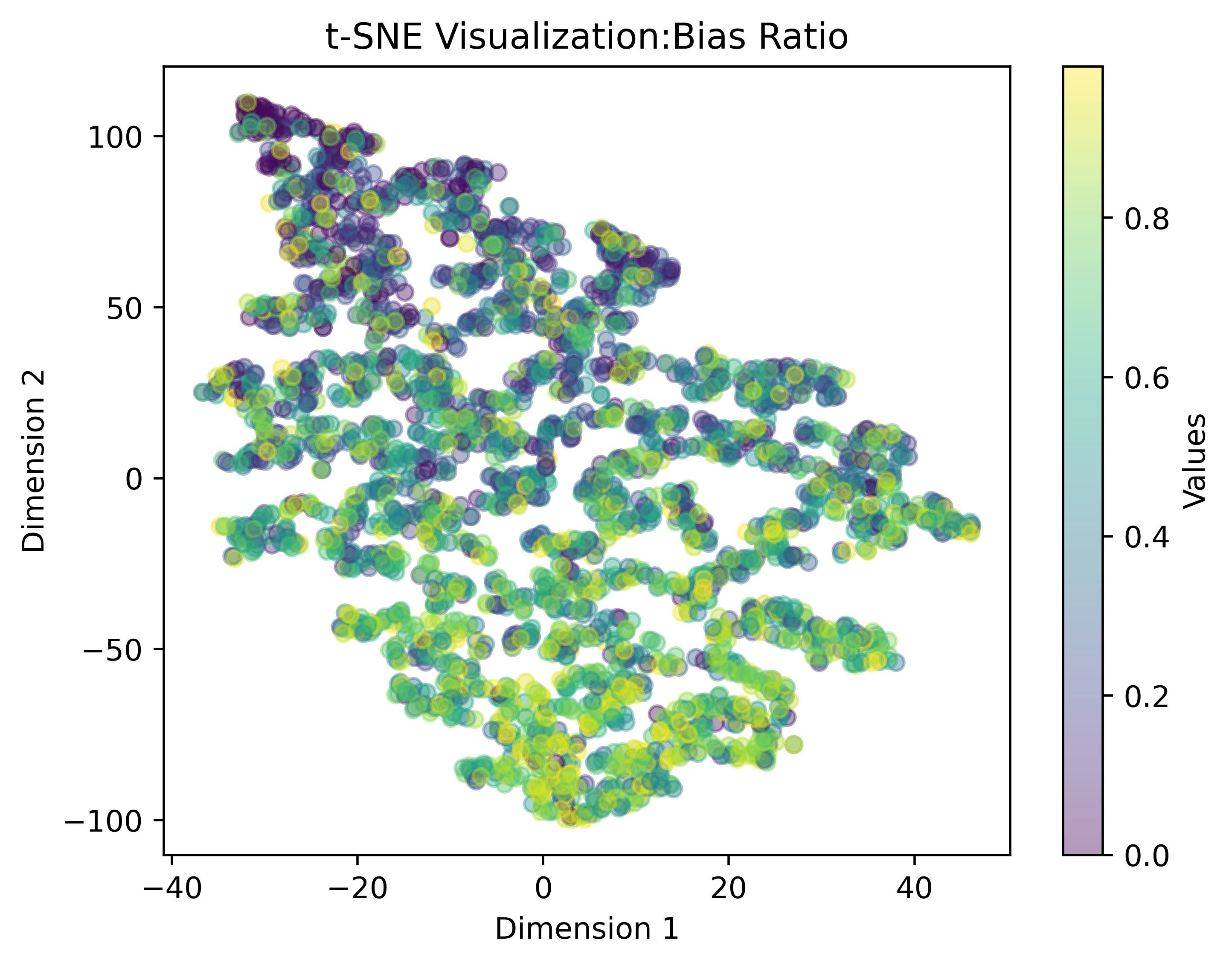}
        \caption{Bias ratio.}
        \label{fig:sub3}
    \end{subfigure}
    \captionsetup{font=normalsize}
    \caption{Risk similarity scatter plot for stocks. This figure illustrates the risk similarity scatter plots for stocks using t-SNE dimensionality reduction of the code embedding output vectors. Each subplot visualizes stock attributes in a two-dimensional vector space, with each point representing a stock. The position of each point is determined by the two-dimensional vectors after dimensionality reduction, and the color represents the attribute value (ranging from 0 to 1). (a) shows the 60-day return standard deviation (STD60) as the color bar, (b) depicts the turnover rate, and (c) illustrates the bias ratio. These visualizations indicate that the code embedding layer can represent stocks with similar risk attributes closer in the vector space, aiding in the understanding of inter-asset risk relationships.}
    \label{fig:sim_tSNE}
\end{figure}

In our analysis of stock return uncertainty, we leveraged t-SNE technology to condense the intricate features of stocks into a comprehensible two-dimensional space. We then employed color coding to visually illustrate the relative standings of stock attribute values, as depicted in \Cref{fig:sim_tSNE}. Specifically, we designated color attributes based on three mainly market indicators: turnover rate, 60-day return standard deviation (STD60), and deviation rate derived from a five-day moving average. The incorporation of these indicators clarifies the closeness of points in relation to these attributes, which facilitates the recognition of visually similar stocks on the visualization chart. 

During our analysis, each attribute of every stock, such as turnover rate and 60-day return standard deviation, underwent transformation into a relative ranking value for visualization purposes. This ranking value was computed by comparing the attribute value of each stock with the corresponding ranked value across the entire stock dataset. Mathematically, this transformation can be expressed as:
\begin{equation}
    \label{eq:vrank}
    c_i = r_i/S,
\end{equation}
where $c_i$ represents the color encoding value for the $i$-th stock, $r_i$ denotes the ranking of that stock on the given attribute, and $S$ stands for the total number of stocks. This calculation method ensures that the color encoding effectively reflects each stock's position relative to all others on the selected attributes, providing an intuitive understanding of attribute similarity in the t-SNE visualization. By integrating this relative ranking formula with t-SNE's dimensionality reduction results, we can aptly capture and depict similarity among stocks across chosen financial indicators.

Upon further analysis, we discovered that revealing clustering patterns among stocks relies not only on spatial distances but also on a comprehensive distribution of attribute similarity. By constructing a similarity network, we unveil the intricate structures underlying these distributions, wherein each node symbolizes a stock, and edge weights are determined by attribute value similarity. This approach enhances the understanding of stock market structure.

\FloatBarrier

\section{Robustness Analysis}
\label{sec:robustness analysis}
We conducted robustness testing on the mixture density network model by training the network 30 times and summarizing the continuous ranked probability score evaluation results for each time. Instead of conducting robustness analysis on GARCH models, we focused on MDN and MDNe models due to the complexity of deep learning models, which are susceptible to variations in predictive performance caused by factors such as initialization and data sample batches. Robustness testing helps ensure the reliability of these models by evaluating their performance consistency across multiple training runs. In contrast, GARCH models typically have simpler structures and fewer parameters, making frequent robustness validation unnecessary.

\begin{figure}
    \centering
    \includegraphics[width=0.7\textwidth]{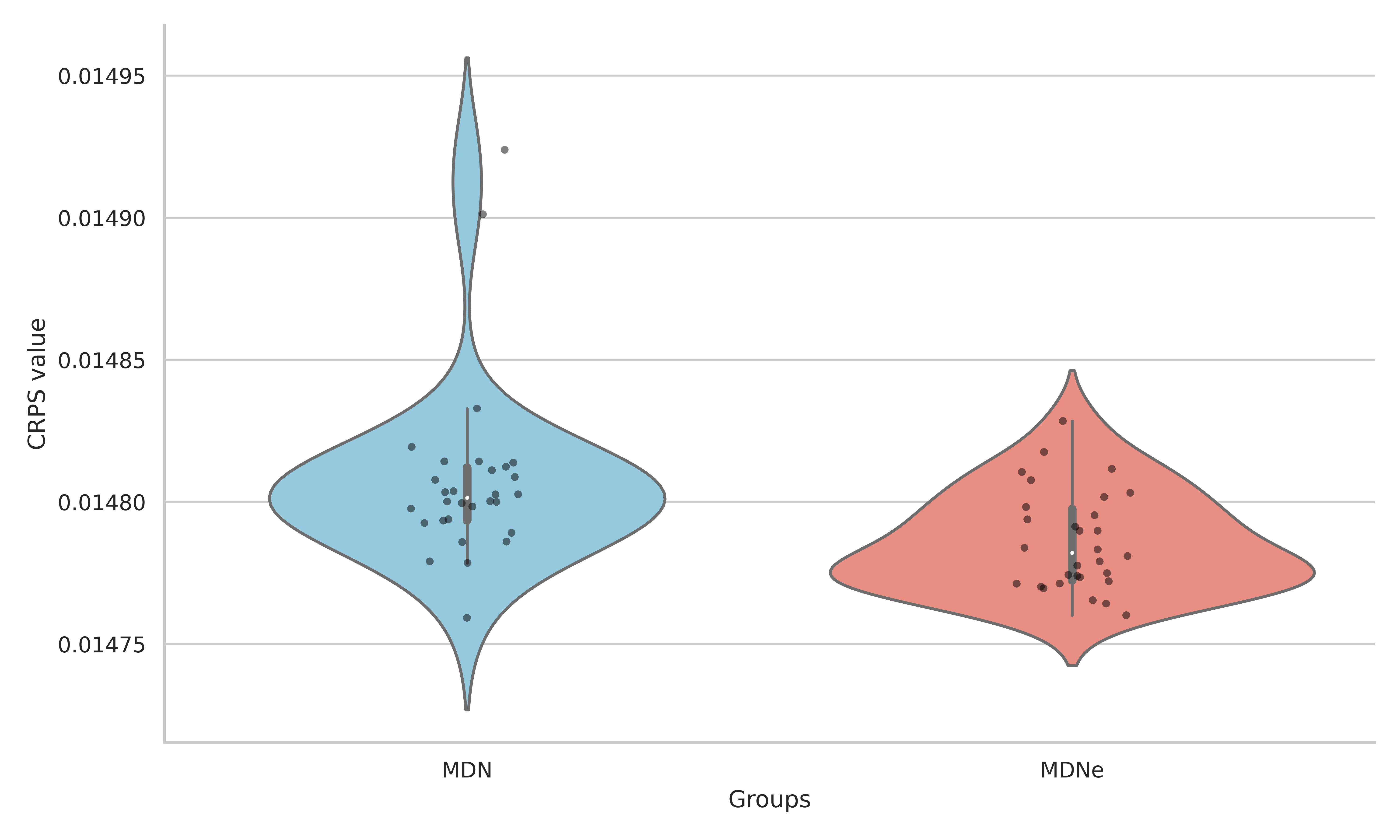}
    \caption{Distribution of CRPS. This figure shows the distribution of CRPS values for the MDN and MDNe models using violin plots. The plots illustrate the spread of CRPS values from 30 different training runs. Each dot within the plot represents an individual training experiment's CRPS value, while the white dot signifies the median. The black lines indicate the interquartile range. The MDNe model displays a more concentrated distribution and a lower median, suggesting higher predictive consistency and accuracy compared to the MDN model.}
    \label{fig:robust}
\end{figure}

The violin plots in \Cref{fig:robust} display the distribution of CRPS values for the MDN and MDNe models. The shape of the violin plot indicates the spread of CRPS values, with MDNe showing a relatively concentrated distribution and a lower median. This suggests that the MDNe model consistently exhibits higher predictive consistency and accuracy across multiple training runs. On the other hand, the CRPS values for the MDN model demonstrate a wider distribution and a slightly higher median, indicating relatively dispersed predictive performance and potentially greater variability across different training instances.

The dots within the violin plots represent the CRPS values for individual training experiments, while the white dots in the center signify the median. By observing the black lines, one can discern the interquartile range, which indicates the spread of the CRPS values. A smaller interquartile range, as seen in the MDNe distribution, suggests that its CRPS values are predominantly concentrated within a narrower range. This typically implies higher stability and reliability in the model's predictions.

To sum up, this figure offers a comparative assessment of the robustness between two models, illustrating that the MDNe model generally demonstrates better consistency and stability in continuous performance evaluation compared to the MDN model. This finding holds practical significance for model selection in forecasting stock market volatility, as we typically favor models that consistently yield reliable results across various conditions.

\FloatBarrier

\section{Conclusion}
\label{sec:conclusion}
This study investigates the prediction of stock market return uncertainty using advanced deep learning technique, specifically the mixture density network (MDN) model. Unlike traditional GARCH models, the MDN model leverages a Gaussian mixture distribution to better capture the complex dynamics and non-traditional features of stock returns. This approach proves particularly effective in accurately describing and predicting market behavior during periods of significant volatility change.
Our empirical analysis demonstrates that the MDN model exhibits robust performance and high predictive accuracy across a wide range of stocks in the Chinese stock market. By deeply exploring the intricate relationships among various stocks and considering multiple financial factors, the MDN model significantly outperforms traditional volatility models. 

In addition to the core model, we introduced the novel application of code embedding and t-SNE visualization technique. These methods allow for mapping the risk similarities among different assets, facilitating the identification of asset clusters with similar risk profiles. This visualization aids portfolio management and risk mitigation strategies, enhancing the practical application of our model in financial decision-making.

Our findings suggest that integrating deep learning with a Gaussian mixture distribution approach improves the accuracy of volatility forecasting and enhances the interpretability and applicability of risk modeling in financial markets. Future research can build on this work by applying these methods to analyze the impact of risk distribution patterns on the market and incorporating additional macroeconomic variables to further refine model performance. Additionally, exploring more advanced deep learning architecture and optimization technique may yield further improvements in predictive accuracy and computational efficiency.

In conclusion, this research introduces a pioneering method that enhances the capability to predict market volatility and understand the uncertainty of stock returns, offering valuable insights and practical tools for financial risk management.

\clearpage
% In the interest of anonymization, please do not include acknowledgements in your submission.
%
%\begin{acks}
%
%	The authors would like to thank Dr. Maura Turolla of Telecom
%	Italia for providing specifications about the application scenario.
%
%	The work is supported by the \grantsponsor{GS501100001809}{National
%		Natural Science Foundation of
%		China}{http://dx.doi.org/10.13039/501100001809} under Grant
%	No.:~\grantnum{GS501100001809}{61273304\_a}
%	and~\grantnum[http://www.nnsf.cn/youngscientsts]{GS501100001809}{Young
%		Scientsts' Support Program}.
%
%
%\end{acks}

% Bibliography
\bibliographystyle{ACM-Reference-Format}
% \bibliography{sample-bibliography}
\bibliography{reference_ec}

% Appendix
\appendix

\end{document}